\documentstyle[12pt,a4,epsfig]{article}

\newcommand{\be}{\begin{equation}}
\newcommand{\ee}{\end{equation}}
\newcommand{\ba}{\begin{eqnarray}}
\newcommand{\ea}{\end{eqnarray}}

\begin{document}

\begin{titlepage}
\begin{flushright}
{BUTP--98/18}\\
\end{flushright}
\vspace{1.2cm}
\begin{center}
{\Large\bf The generating functional for hadronic weak \\ [0.3cm]
interactions and its quenched approximation}\\
\vspace{2.5cm}
{\bf Elisabetta Pallante\footnote{e--mail:
     pallante@itp.unibe.ch}  } 
\\[0.6cm]
 {\em University of Bern, Sidlerstrasse 5,\\
CH--3012 Bern, Switzerland}\\[0.5cm]
\end{center}
\vfill
\vspace{0.6cm}
\begin{abstract}

We derive the generating functional of $\vert\Delta S\vert =1,\, 2$
hadronic weak interactions at one loop 
for a generic number of flavours and its counterpart in the 
quenched approximation. A systematic analysis of the ultraviolet divergences
in the full theory (with and without a singlet dynamical field) and in 
the quenched case is performed.
We show that the quenched chiral logarithms in the presence of
 weak interactions amount to a redefinition of the weak mass term in
 the $\Delta S=\pm 1$ weak effective Lagrangian at leading order.
Finally, we apply the results to $B_K$ and $K\to\pi\pi$ matrix elements with
$\Delta I=1/2,3/2$ to analyze the modifications induced by quenching 
on the coefficients of chiral logarithms in the one-loop corrections.

\end{abstract}
\vspace{1cm}
\hspace{0.6cm}{\small PACS: 12.39.Fe, 12.38.Gc, 13.25.Es}
\vspace{0.5cm}

\vspace*{1cm}
\setcounter{equation}{0}
\setcounter{figure}{0}

\vfill 
\end{titlepage}

\renewcommand{\theequation}{\arabic{section}.\arabic{equation}}
\setcounter{equation}{0}
\section{Introduction}

Nonleptonic weak interactions of light mesons are still one of the main
unresolved issues of the Standard Model (SM). Two major aspects still await
for a theoretical understanding: the $\Delta I=1/2$ rule
and the mechanism which mediates CP violation. 
Also, interaction mechanisms beyond the SM find a
rich spectrum of implications in this context.

The short-distance QCD description of the operators that mediate nonleptonic
weak transitions has been clearly formulated \cite{QCDSD}. QCD asymptotic
freedom and Operator Product Expansion are used to integrate out heavy 
degrees of freedom of the SM Lagrangian
down to $\mu < m_c$. At this scale nonleptonic weak transitions are mediated
by effective four-quark operators of light quarks u, d, s. The QCD evolution
of their Wilson coefficients is determined by RG equations.

The evaluation of the matrix elements of the four-quark effective operators
between light meson states is instead a pure non perturbative problem.
It involves the knowledge of QCD contributions at long distances, namely 
from $\mu \sim m_c$ down to scales $\mu\simeq \Lambda_{QCD}$ or the typical
light meson mass. 

At long distances two approaches are viable. The first is 
the implementation of light hadrons matrix elements on the lattice,
 the second
is the use of an effective Lagrangian for weak interactions
where the dynamical degrees of freedom are the light mesons. 
The weak Lagrangian can be formulated on the same principles 
as the Chiral Perturbation Theory (ChPT)
Lagrangian for strong interactions. It is an expansion in powers of the
momenta of light mesons and light quark masses.
The leading order weak Lagrangian is known since a long time 
\cite{LOWEAK,Bernard}, 
while the derivation of the next-to-leading order 
Lagrangian was done in Refs. \cite{KMW,EKW} for $N=3$ flavours.
This second approach is based on symmetry principles. Its 
 advantage is that it allows for a perturbative treatment of weak processes at
 long distances. Its major drawback is the lack of predictivity in the
 parameters, residues of short-distance physics,
 which regulate the various interactions.

The computation of weak matrix elements on the lattice originates
from {\em first principles} and in this sense it gives the correct answer 
to the problem. On the other hand it still suffers from the presence of major
sources of systematic errors: finite volume effects, unphysical quark masses,
 the need of computing {\em unphysical} matrix elements \cite{BERNARD}
 (see also Ref. \cite{MARTI} for recent alternative proposals), 
in order to maximally avoid mixing with lower dimension operators
 in decay processes with two or more particles in the final state, and
finally the fact that still most of the lattice evaluations are done in the
quenched approximation. In this work we focus on formal aspects of the 
 last problem within the ChPT framework, while in the immediate 
phenomenological applications we shall explore in a quantitative way
 the deviations of the quenched weak matrix elements from the physical ones. 
Further results on this last argument will be reported in Ref. \cite{E2}.

Our main scope is to provide a systematic treatment of the quenched
approximation for nonleptonic weak interactions. On the other hand, the same
approach will allow to add a few new results to previous analyses 
of weak interactions within the full (unquenched) theory.
The approach that we adopt is
the one of the generating functional within the framework of the
low energy effective Lagrangian.
This approach was applied in Refs. \cite{pl,npb} 
to the quenched approximation of strong interactions (also known as quenched
ChPT \cite{qCHPT}). 
The first derivation of the generating functional of standard (unquenched) 
ChPT can be found in Refs. \cite{gl84,gl85}. The analogous derivation in the
weak sector can be found in Refs. \cite{KMW,EKW}, where the
generating functional of hadronic weak interactions has been derived 
in the full theory with $N=3$ flavours.

In this work we calculate the ultraviolet divergences of the weak generating
functional at one loop in the full theory and in the 
quenched approximation. This involves, as a first step,
 the derivation of the generating functional in the full theory with a generic
 number of flavours $N$ and in the presence of a singlet dynamical field.
The flavour number dependence was not known before, while the inclusion of the
singlet dynamical field was sketched in Ref. \cite{EKW} (see Appendix A
thereof) in the analysis of the $(8_L,1_R)$ sector.
Here we introduce the singlet dynamical field at first place and derive the
modifications induced both in the $(8_L,1_R)$ and the  $(27_L,1_R)$ operators
which carry the ultraviolet divergences.
As an outcome, 
we construct the minimal basis of ultraviolet divergent
effective meson operators which appear at next-to-leading order 
both in the octet and 27-plet weak Lagrangians.
We limit the present analysis to the case of degenerate light quark masses. 
In this case, once the ultraviolet divergences of the generating functional 
are known in the
full and in the quenched theory, all the coefficients of chiral logarithms are
known in both theories, for any nonleptonic weak matrix 
element of light mesons. For non degenerate light quark masses, also 
ultraviolet finite logarithms of the type $\log (m_K^2/m_\pi^2)$ do appear.

The whole derivation is done at infinite volume, with
the intention of clarifying some of the formal aspects of the quenched
approximation to weak interactions. The infrared behaviour of the quenched
approximation in the presence of weak interactions is not considered here. 
However, we anticipate that its analysis with
the use of the generating functional follows the same lines of the analysis
done in the case of purely strong interactions in Ref. \cite{npb}.

The next question to answer is how much the quenched approximation modifies
the predictions of weak matrix elements. Knowing the value of all the chiral
logarithms in the full theory and the corresponding one in the quenched
approximation, one can at least quantitatively predict how much quenching
modifies their contribution to weak matrix elements. Here, we focus on the
analysis of $B_K$ and $K\to\pi\pi$ matrix elements. In the last case quenching
effects turn out to be of considerable size. Also, their pattern
in respect to the $\Delta I=1/2$ rule can be clarified:
 while the full ChPT one loop corrections tend
to support the enhancement of the $\Delta I=1/2$ amplitude, quenched
modifications (at infinite volume) tend to suppress the $\Delta I=1/2$
dominance. 
The modifications induced by quenching, together with the analysis of
{\em unphysical} $K\to\pi\pi$ matrix elements as they are implemented on the
lattice are further considered in Ref. \cite{E2}.  

The plan of the paper is as follows. In Section \ref{QLAG} the effective
Lagrangian for hadronic weak interactions with $\vert\Delta S\vert =1,\, 2$
 is extended to its graded version, that reproduces the quenched
approximation within Chiral Perturbation Theory. In Section \ref{WGF} the
ultraviolet divergences of the generating functional for hadronic weak 
interactions at one loop 
are derived in the full theory for a generic number of flavours and in the 
quenched approximation. A peculiar behaviour of the quenched approximation 
in the presence of weak interactions is the generation of quenched chiral
logarithms in addition to the ones produced by strong interactions \cite{npb}.
How they can be formally interpreted as the rescaling of a 
parameter in the leading order weak Lagrangian is clarified in Section 
\ref{SingInt}.
We treat separately the bosonic contribution to the generating functional 
in subsections \ref{BOSSECT}, \ref{BOSNSING}, \ref{SingInt} and \ref{BOSCR},
and the fermionic contribution in subsection \ref{FER}.
The reader who is not interested in the technical details
of the derivation can skip the full section \ref{WGF}.
In Section \ref{CR} we give the final result for the ultraviolet divergences 
in the quenched approximation, while in subsection \ref{DVCT} we define the
divergent counterterms in the full theory and compare with their quenched
counterpart (see Tables (\ref{TABCTR8}) and
 (\ref{TABCTR27}) for the comparison).
 This analysis is basically useful for the 
phenomenological applications considered in Section \ref{phen}. Here we focus
on the contribution to weak observables coming from the chiral logarithms and
how they are modified by quenching. The analysis is performed at infinite
volume. 
In particular, in subsection \ref{BKpar} we rederive known results for the
$B_K$ parameter, with the aim of clarifying  the structure of the counterterms
and their flavour number dependence. In subsection \ref{kppMAT} we 
analyze $K\to\pi\pi$ matrix elements with $\Delta I=1/2$ and $3/2$; we
consider the full matrix elements and their quenched approximation.
 Since we work in the continuum at
infinite volume, we give all the predictions in Minkowski space-time.
We conclude in Section \ref{CONC}. There are two appendices. In Appendix
\ref{LIST} the list of divergent counterterms in the octet and 27-plet case is
given and in the presence of the singlet field. In Appendix \ref{SINGLET} 
the single contributions to the bosonic determinant in the singlet sector are
illustrated.  

\renewcommand{\theequation}{\arabic{section}.\arabic{equation}}
\setcounter{equation}{0}
\section{The quenched Lagrangian for weak interactions}

\label{QLAG}
The complete quenched ChPT Lagrangian for light pseudoscalar mesons 
at leading order in the chiral expansion (i.e. at order $p^2$ and linear in 
the light quark masses) can be written as follows:
\be
{\cal L} = {\cal L}_{\Delta S=0}+ {\cal L}_{\Delta S= 1}+
{\cal L}_{\Delta S=2}\, .
\label{totL}
\ee
The Lagrangian ${\cal L}_{\Delta S=0}$ is the quenched version of the leading 
order Lagrangian for strong interactions \cite{npb,qCHPT}
\ba
{\cal L}_{\Delta S=0}
&=&V_1(\Phi_0) \mbox{str} (u_{\mu s}u^\mu_s )+V_2(\Phi_0) 
\mbox{str} (\chi_{+s} )-V_0(\Phi_0) \nonumber\\
&&+V_5(\Phi_0) D_\mu\Phi_0 D^\mu \Phi_0 \;  .
\label{Ls}
\ea
It is invariant under the graded chiral
symmetry group $\left[ SU(N|N)_L\otimes SU(N|N)_R \right] \odot U(1)_V$ with
$N$ physical flavours and $N$ ghost flavours.
The graded fields\footnote{The quenched counterpart of a standard CHPT
  quantity is either denoted with capital letters (as in $\phi\to \Phi$) or
  with the $s$ subscript} are defined with the usual notation \cite{npb}
\ba
u_{\mu s}&=& iu_s^\dagger D_\mu U_s u_s^\dagger = u_{\mu s}^\dagger \nonumber\\
\chi_{+s}&=&u_s^\dagger\chi_s u_s^\dagger +u_s\chi_s^\dagger u_s\, ,
\label{FIELDS}
\ea
where $U_s = u_s^2$ is the exponential representation of the graded meson 
field: 
\[
U_s = \exp (\sqrt{2} i\, \Phi/F) \; \; ,
\]
$F$ is the bare quenched pion decay constant (with renormalized value $F_\pi =
93$ MeV) and $\Phi$ is a 
hermitian non traceless $2\times 2$ block matrix
\[
\Phi = \left( 
\begin{array}{cc}
\phi & \theta^\dagger  \\
\theta & \tilde\phi
\end{array}  \right) \; , ~~~\mbox{str}(\Phi) = \Phi_0 =
\phi_0-\tilde\phi_0 \; \; .
\]
The potentials $V_i (\Phi_0)$ in Eq. (\ref{Ls}) are real and even
functions of the super-$\eta^\prime$ field $\Phi_0 = \mbox{str}(\Phi)$.

The Lagrangians ${\cal L}_{{\tiny{\Delta S= 1}}}$ and 
${\cal L}_{{\tiny{\Delta S=2}}}$ in Eq. (\ref{totL}) are the quenched version 
of the leading order Lagrangians which 
 mediate strangeness changing non leptonic weak interactions in one and 
two units. The quenched
version of the weak effective Lagrangians is the generalization to a 
graded group (i.e. with $N$ physical flavours and $N$ ghost flavours)
of the leading order weak effective Lagrangians given in 
Refs. \cite{Bernard,KMW,EKW}, with the inclusion of a singlet dynamical field. 
Using the same notation as for the strong Lagrangian, 
they can be written as follows:
\ba
{\cal L}_{{\tiny{\Delta S=1}}}
&=&\tilde{V}_8(\Phi_0) \mbox{str}(\Delta_{s 32}u_{\mu s}u^\mu_s )
+\tilde{V}_5(\Phi_0)\mbox{str}(\Delta_{s 32}\chi_{+s} )
+\tilde{V}_0(\Phi_0)\mbox{str}(\Delta_{s 32}u_{\mu s}) \mbox{str}(u^\mu_s )
\nonumber\\
&& + {\mbox{h.c}} 
+ \tilde{V}_{27}(\Phi_0)t^{ij,kl}\mbox{str}(\Delta_{s ij}u_{\mu s})
\mbox{str}(\Delta_{s kl}u^\mu_s) 
\label{WEAK}
\ea
and 
\be
{\cal L}_{{\tiny{\Delta S=2}}}=\tilde{V}_{27}^{\tiny{\Delta S=2}}(\Phi_0)\, 
t^{ij,kl}\mbox{str}(\Delta_{s ij}u_{\mu s})
\mbox{str}(\Delta_{s kl}u^\mu_s)\, ,
\label{DS2}
\ee
where the potentials $\tilde{V}_i(\Phi_0)$ are again real and even
functions of the super-$\eta^\prime$ field $\Phi_0$.
The $\Delta S=1$ Lagrangian in Eq. (\ref{WEAK}) contains the octet $(8_L, 1_R)$
operators in the first line and the 27-plet $(27_L, 1_R)$ operator in the 
second line. The octet term associated to the potential 
$\tilde{V}_0(\Phi_0)$ is induced by the inclusion of the singlet dynamical 
field.

The operator $\Delta_{s ij}$ is the graded realization of the usual 
projection matrix onto the octet components of the chiral fields. 
It is given by 
$ \Delta_{s ij}=u_s \lambda_{ij} {1+\tau_3\over 2}u_s^\dagger$, with
$\left (\lambda_{ij}\right )_{ab} = \delta_{ia}\delta_{jb}$.
The tensor $t^{ij,kl}$ in Eqs. (\ref{WEAK}), (\ref{DS2}) projects onto the
$(27_L,1_R)$ component of the interacting fields, with $\vert\Delta S\vert
=1$ or 2. It is a completely symmetric tensor that satisfies 
\be
t^{ij,kl}=t^{kl,ij}
\label{tsym}
\ee
 and the following trace zero conditions 
\be
\sum_{i}\, t^{ii,kl}=0,~~~~~~~~~\sum_{i}\, t^{ij,ki}=0\, .
\label{tPROP}
\ee
Note that in the fundamental (unquenched) theory the first condition in Eq. 
(\ref{tPROP}) guarantees that only the traceless component of any  
operators ${\cal O}_1,\,{\cal O}_2 $ contributes to the  $(27_L,1_R)$ term
$ t^{ij,kl}\mbox{tr}(\lambda_{ij}{\cal O}_1 )
\mbox{tr}(\lambda_{kl}{\cal O}_2 )$, with  
$\left (\lambda_{ij}\right )_{ab} = \delta_{ia}\delta_{jb}$. In the quenched 
case the same condition  guarantees that  only the traceless part of the 
$(1,1)$ component of the graded operators  ${\cal O}_{s 1},\,{\cal O}_{s 2}$
 contributes to the  $(27_L,1_R)$ term 
$ t^{ij,kl}\mbox{str}(\lambda_{ij}{1+\tau_3\over 2}{\cal O}_{s 1} )
\mbox{str}(\lambda_{kl}{1+\tau_3\over 2}{\cal O}_{s 2} )$.

In the $\Delta S=\pm 1$ case and in $SU(3)$ the $ t^{ij,kl}$ tensor
 has the following components
\ba
&&t^{21,13}=t^{13,21}=t^{12,31}=t^{31,12}={1\over 3} \nonumber\\
&&t^{22,23}= t^{23,22}=t^{22,32}=t^{32,22}=-{1\over 6} \nonumber\\
&&t^{33,23}= t^{23,33}=t^{33,32}=t^{32,33}=-{1\over 6} \nonumber\\
&&t^{11,23}= t^{23,11}=t^{11,32}=t^{32,11}={1\over 3}
\label{T27}
\ea
and $0$ otherwise, while for  $\Delta S=2$ interactions it is given 
by\footnote{In the
      generalization to $N>3$ flavours the same tensor as in Eqs. (\ref{T27})
      and (\ref{T272}) is used.} 
\be
t^{23,23}=t^{32,32}=1,\,\,\,  0\, {\mbox{otherwise}}\, .
\label{T272}
\ee
The weak Lagrangians in Eqs. (\ref{WEAK}) and (\ref{DS2}) 
are CPS invariant \cite{Bernard}, CP even and S even, where the 
 S (``switching'')
invariance denotes the invariance under the interchange of $2$ and $3$ 
components ($s\leftrightarrow d$).
One can simply obtain the CP odd, S odd $\Delta S =1$ Lagrangian from Eq. 
(\ref{WEAK}) by replacing everywhere $\tilde{V}_i\to i \tilde{V}_i^- $, 
$\Delta_{s32}\to -  \Delta_{s32}$ in the octet terms and
assigning
 the opposite sign to $t^{32,ii},\, t^{ii,32},\,t^{21,13},\, t^{13,21}$ in
the 27-plet terms.

The symmetry under the graded group is made local in the presence of 
graded external sources $l^\mu_s, r^\mu_s, s_s, p_s$. The covariant
derivative over the field $U_s$ is defined as $D^\mu U_s=\partial^\mu
U_s-ir^\mu_s U_s+iU_sl^\mu_s$ and the field $\chi_s=2B_0(s_s+ip_s)$ of Eq. 
(\ref{FIELDS}) contains
the external scalar ($s_s$) and pseudoscalar ($p_s$) sources. Since we are not
interested in studying matrix elements with  spurious fields as external legs
we put to zero all spurious external sources. With this reduction a generic
graded source reads
\[
j_s = 
\left(
\begin{array}{cc} 
j  & 0 \\
0 & 0
\end{array} 
\right) \; ,\; j= p,v_\mu, a_\mu \; \; ,
\]
and the scalar source contains the quark mass matrix ${\cal M}$ at the leading
order
\[  
s_s=\left(  
\begin{array}{cc} 
{\cal M } + \delta s  & 0  \\
0 & {\cal M} 
\end{array}  \right)  \; \; .
\] 
In what follows the quark mass matrix will be taken
proportional to the unit matrix: ${\cal M} = m_q { \bf 1}$. 
 
All the potentials $V_i(\Phi_0)$ and $\tilde{V}_i(\Phi_0)$ in
Eqs. (\ref{Ls}, \ref{WEAK}, \ref{DS2}) 
can be expanded in powers of  $\Phi_0^2$. 
We define the first terms in the expansion as follows 
\ba
V_0(\Phi_0) &=& \frac{m_0^2}{2N_c} \Phi_0^2 +O(\Phi_0^4) \; \; ,
\nonumber \\ 
V_{1,2}(\Phi_0) &=& \frac{F^2}{4}+{1\over 2}v_{1,2}\,\Phi_0^2 +
O(\Phi_0^4) \; \; , \nonumber \\
V_5(\Phi_0) &=& \frac{\alpha}{2N_c} + O(\Phi_0^2) \; \; ,
\ea
in the strong sector, where $m_0^2$ is the squared singlet mass and $\alpha$
 is a new parameter associated with the kinetic term of the singlet field, 
\ba
\tilde{V}_8(\Phi_0)&=& g_8+ {1\over 2}\tilde{v}_8\,\Phi_0^2 +
O(\Phi_0^4) \; \; , \nonumber \\
\tilde{V}_5(\Phi_0)&=& g_8^\prime + {1\over 2}\tilde{v}_5\,\Phi_0^2 +
O(\Phi_0^4) \; \; , \nonumber \\
\tilde{V}_{27}(\Phi_0)&=& g_{27}+ {1\over 2}\tilde{v}_{27}\,\Phi_0^2 +
O(\Phi_0^4) \; \; , \nonumber \\
\tilde{V}_0(\Phi_0)&=& \overline{g}_8+ {1\over 2}\tilde{v}_0\,\Phi_0^2 +
O(\Phi_0^4) \; \; 
\ea
in the $\vert\Delta S\vert =1$ weak sector and 
\ba
\tilde{V}_{27}^{\tiny{\Delta S=2}}(\Phi_0)&=& g_{27}^{\tiny{\Delta S=2}}
+ {1\over 2}\tilde{v}_{27}^{\tiny{\Delta S=2}}\,\Phi_0^2 +O(\Phi_0^4) 
\ea
in the $\Delta S=2$ weak sector.
The weak couplings $g_8,\, g_{27},\, g^\prime_8,\, \bar{g}_8$ 
and $g_{27}^{\tiny{\Delta
    S=2}}$  can be written in terms of dimensionless couplings as follows
\ba
&&\hspace{-0.8truecm}g_8=CF^4G_8\, ,~~~~ g_{27}=CF^4G_{27}\,
,~~~~g^\prime_8=CF^4G_8^\prime\, ,~~~~\bar{g}_8=CF^4\bar{G}_8\, ,\nonumber\\
&&\hspace{-0.8truecm}g_{27}^{\tiny{\Delta S=2}}=
C^{\tiny{\Delta S=2}} F^4G_{27}\, ,
\label{GWEAK}
\ea  
where the constant $C$ contains the CKM matrix elements via
\be
C=-{3\over 5}{G_F\over\sqrt{2}}V_{ud}V_{us}^*\, ,
\ee
 while the constant $C^{\tiny{\Delta S=2}}$ can be written as follows
\be
C^{\tiny{\Delta S=2}}=-{G_F\over 4}{\cal F} ( m_t^2,\, m_c^2,\, M_W^2,\,
{\bf{V}}_{CKM} )\, ,
\label{CDS2}
\ee
where ${\cal F} ( m_t^2,\, m_c^2,\, M_W^2,\,
V_{CKM} )$ is a known function of the $W$
boson mass, the heavy quark masses and CKM matrix elements \cite{BURAS}.
In the large-$N_c$ limit the dimensionless couplings $G_8$ and $G_{27}$ 
are $G_8=G_{27}=1$.
Notice that the 27-plet couplings $g_{27}$ and $g_{27}^{\tiny{\Delta S=2}}$
are the same, modulo the weak-sector coefficients $C$ and 
$C^{\tiny{\Delta S=2}}$. We expect that the same is true for the full 
potentials $\tilde{V}_{27}$ and $\tilde{V}_{27}^{\tiny{\Delta S=2}}$, so that 
$\tilde{V}_{27}^{\tiny{\Delta S=2}} =(C^{\tiny{\Delta S=2}}/C) \,
 \tilde{V}_{27}$.

\renewcommand{\theequation}{\arabic{section}.\arabic{equation}}
\setcounter{equation}{0}
\section{The Weak generating functional to one loop}

\label{WGF}

The derivation of the generating functional for hadronic weak interactions and
its counterpart in the quenched approximation can be done following the same
lines of Ref. \cite{npb}, where the quenched generating functional for low
energy strong interactions has been derived within the framework of ChPT. 
We expand the leading order action around the classical solution
up to and including quadratic fluctuations. We write the field $U_s$ as:
\[
U_s= u_s ~e^{i \Xi}~ u_s \; \; ,
\]
where $\bar{U_s}=u_s^2 $ is the classical solution to the
equations of motion. In the absence of spurious external sources it
reduces to 
\[  
u_s=\left(  
\begin{array}{cc} 
{ u }   & 0  \\
0 & {\bf 1} 
\end{array}  \right)  \; \; .
\]
We decompose the fluctuation $\Xi$ similarly to the field $\Phi$ and write:
\[
\Xi = \left( 
\begin{array}{cc}
\xi & \zeta^\dagger  \\
\zeta & \tilde\xi
\end{array}  \right) \;\; , \; \; \; \; \mbox{str}(\Xi)=
\sqrt{N}(\xi_0-\tilde\xi_0) \; \; .
\]
Each field can be decomposed in terms of the generalized 
$ \hat\lambda_a$ matrices  
\be 
\xi =\sum_{a=0}^{N^2-1} \hat\lambda_a \xi^a , 
\ee
where $\hat\lambda_a =\lambda_a/\sqrt{2}, {\bf{1}}/\sqrt{N}$ for $a=1,\ldots
N^2-1$ and $a=0$ respectively. It is useful to define a graded vector  for
the bosonic fields as follows: $X^T= (\xi^T ,\tilde\xi^T)$ and
$\xi^T=(\xi^0, \xi^1,\xi^2,\ldots ,\xi^{N^2-1})$, the same for $\tilde\xi^T$. 
The fermionic ghost fields are collected in the vector 
$\zeta^\dagger = (\zeta^{\dagger 0}, \zeta^{\dagger
    1},\ldots ,  \zeta^{\dagger N^2-1})$. With this notation 
the complete action for strong and weak interactions up to quadratic
fluctuations can be written in a compact form\footnote{As we did in the strong
  case we disregard the infinite chain of terms containing powers of the
  super-$\eta^\prime$ field at the classical solution $\bar{\Phi}_0$, coming
  from the expansion of the potentials $V_i(\Phi_0)$ and
  $\tilde{V}_i(\Phi_0)$.}
\be
S[\Phi]= S[\bar\Phi ]-{F^2\over 4}\int~dx\left\{ X^T D X
+2 \zeta^\dagger D_\zeta \zeta \right\} +O(\Xi^3)\, ,
\label{S1}
\ee
where summation over flavour indices is implicit. Notice also that the bosonic
differential operator $D$ is a graded $2\times 2$ matrix acting on $X$.
Using the action (\ref{S1}), the quenched generating functional to one loop
can be formally written as follows
\be
e^{iZ^{\mbox{\tiny{q}}}_{\mbox{\tiny{1\,loop}}}}= {\cal N}{\det{D_\zeta}\over 
(\det{D})^{1\over 2}}\, .
\label{DET1}
\ee
In the following subsections we treat separately the bosonic part and 
the fermionic ghost part of the generating functional in Eq. (\ref{DET1}).

\subsection{The bosonic sector}

\label{BOSSECT}

While in the purely strong case treated in Ref. \cite{npb} a mixing is 
only induced amongst the  singlet 
components of the  physical $\xi$ and the ghost $\tilde{\xi}$ 
fluctuation field, 
weak interactions induce a mixing also in the non singlet sector. For
this reason a compact representation of the generating functional in the
bosonic sector in terms of graded vectors and graded operators turns out to be
convenient.
The differential operator $D$ in Eq. (\ref{S1}) is a $2\times 2$ graded
differential matrix acting on the graded vector $X$. It can be written as
follows
\be
D_{ab} = d_{\mu s} G_{ab} d^\mu_s +F_{ab}\, ,
\label{BOSD}
\ee
where $d^\mu_s$ is the graded covariant derivative given by
\be
d_{\mu s}=d_\mu {1+\tau_3\over 2}+\tilde{d}_\mu{1-\tau_3\over 2}\, ,
\label{COV}
\ee 
with $d_\mu$ the covariant derivative acting on $\xi$ and
$\tilde{d}_\mu$ the covariant derivative acting on $\tilde\xi$. In the absence
of spurious external sources $\tilde{v}_\mu , \tilde{a}_\mu =0$ 
the latter reduces to the partial derivative $\tilde{d}_\mu = \partial_\mu$. 
The graded operator $G_{ab}$ is given by 
\be
G_{ab}=\delta_{ab}\tau_3+\alpha_{ab}\, 
\label{CONN}
\ee
and it is hermitian. The operators $F_{ab}$ in Eq. (\ref{BOSD}) 
and $\alpha_{ab}$ in Eq. (\ref{CONN}) are in turn $2\times 2$ graded matrices. 
The term $\alpha_{ab}$ is induced by weak interactions and we call it the weak
connection in analogy with the connection tensor which appears in the metric
of a curved space.  To compute the ultraviolet divergences of the generating
functional to one loop with standard techniques it is appropriate to rescale
the coefficient of the d'Alambertian in Eq. (\ref{BOSD}) to one. 
We generalize the rescaling procedure used in Ref. \cite{KMW} to
the quenched case and in the presence of a singlet component.
We define the hermitian operator $G$ as $G=gg=\bar{g}^T\tau_3\bar{g}$, where 
$\bar{g}_{ab}=\sqrt{\delta_{ab}+\tau_3\alpha_{ab}}$ and $\bar{g}^T_{ab}=
\sqrt{\delta_{ab}+\alpha_{ab}\tau_3}$. By commuting $g$ with the covariant
derivative in Eq. (\ref{BOSD}) 
it is easy to show that the following identity holds
\be
d_s ggd_s = gd_sd_sg+ [d_s,g]d_sg-gd_s[d_s,g]-[d_s,g][d_s,g]\, .
\ee
With the use of the $\bar{g}$ definition the quadratic fluctuation of the
bosonic Lagrangian can be rewritten as $X^{\prime T}D^\prime X^\prime$, where
the rescaled differential operator is now given by 
\be
 D^\prime =
 d_s^2\tau_3+\bar{g}^{T{}^{-1}}\,F\,\bar{g}^{{}^{-1}}
+\bar{g}^{T{}^{-1}} \left ( [d_s,g]d_sg-gd_s[d_s,g]-[d_s,g][d_s,g]\right )
\bar{g}^{{}^{-1}} \, ,
\label{DEFRES}
\ee
where flavour indices have been omitted for simplicity, 
and the rescaled fluctuation field is defined as 
$X^\prime = \bar{g}X, X^{\prime {}^T} = X^T\bar{g}^T $.
This is valid provided that $\bar{g}^{-1}$ does exist, that is our case.
The rescaled operator $D^\prime$ has now the treatable form with unit 
coefficient in the double derivative term. In terms of the rescaled
fluctuation fields the bosonic part of the 
generating functional is now given by
\be
e^{iZ_b}= {\cal N}e^{i\overline{Z}_b}\int dX^{\prime {}^T}dX^\prime~
(\det{G\tau_3})^{-{1\over 2}}~e^{-i{F^2\over 4}\int dx\, 
X^{\prime{}^T} D^\prime  
X^\prime}\, ,
\label{BOSGF}
\ee
where $\overline{Z}_b$ is the classical contribution to the generating
functional and  we made use of the identity
$\det{(\bar{g}^T\bar{g})}=\det{(gg\tau_3)}=\det{(G\tau_3)}$.
The complete bosonic determinant is now given by $(\det{D})^{-1/2}
=(\det{D^\prime })^{-1/2}(\det{G\tau_3})^{-1/2}$.
 
As it is the relevant
one, we limit the analysis of the generating functional 
to the first order in the expansion in powers of the
weak interaction coupling $G_F$. The weak connection $\alpha_{ab}$ in
Eq. (\ref{CONN}) is of
order $G_F$, while the operator $F_{ab}$ contains both strong interaction
terms and order $G_F$ ones.
Its explicit expression can be written as follows:
\be
F_{ab} = -{1\over 2}\{ N^-_{ab},d_s\}+{1\over
  2}[d_s,N^+_{ab}]+\hat{\sigma}_{ab} +\hat{\omega}_{ab}\, ,
\label{DEFF}
\ee
where $\hat{\sigma}_{ab}$ is the usual graded strong interaction operator as
defined in Ref. \cite{npb}, 
while $N^-_{ab},\,  N^+_{ab}$ and $\hat{\omega}_{ab}$ are 
graded weak interaction operators.

Expanding $g$ and $\bar{g}^{-1}$ in powers of the weak connection 
$\alpha_{ab}$ (i.e. in powers of $G_F$)
in Eq. (\ref{DEFRES}) and keeping up to order $G_F$ terms, the rescaled 
bosonic differential operator can be written as follows
\be
D^\prime_{ab}=d_s^2\delta_{ab}\tau_3 + F_{ab} -{1\over 2} \left
  (\alpha_{al}\tau_3\hat{\sigma}_{lb} + \hat{\sigma}_{al}\tau_3 \alpha_{lb}
\right )-{1\over 2}[d_s,[d_s,\alpha_{ab} ]] +O(G_F^2)\, ,
\ee
with $F_{ab}$ given in Eq. (\ref{DEFF}).

The operators $F_{ab}$ and $\alpha_{ab}$ act on various subspaces of the
graded space. They can be classified as follows: 
 1) the physical subspace of the fluctuation field $\xi$ 
 with projection operator  $(1+\tau_3)/2$, 
2) the bosonic ghost subspace of the fluctuation field $\tilde\xi$ 
 with projection operator $(1-\tau_3)/2$, 
3) the mixed physical--bosonic ghost subspace with projection operator 
$(1-\tau_1)$, or $[(1+\tau_3)/2-(\tau_1+i\tau_2)/2]$ and its transposed.
For this reason 
the weak connection $\alpha_{ab}$ admits the following decomposition into the
four projected components:
\be
\alpha_{ab}= \alpha^{11}_{ab}{1+\tau_3\over 2}
+\alpha^{22}_{ab}{1-\tau_3\over 2}+ \alpha^{12}_{ab}{\tau_1+i\tau_2\over 2}
+ \alpha^{21}_{ab}{\tau_1-i\tau_2\over 2}\, , 
\label{DEC}
\ee
where\footnote{The notation $\langle\ldots\rangle$ stands everywhere for the
trace over flavour indices.}
\ba
 \alpha^{11}_{ab}&=&  {1\over 2} g_8 k \langle\Delta\{\hat\lambda_a,\hat
\lambda_b\}\rangle
+g_{27}q\langle\Delta \hat\lambda_a\rangle\langle\Delta \hat\lambda_b\rangle +
{1\over 2}\bar{g}_8\sqrt{N}k\left (\langle\Delta\hat\lambda_a\rangle
\delta_{0b} +\langle\Delta\hat\lambda_b\rangle\delta_{0a}\right) \nonumber\\
\alpha^{22}_{ab}&=& 0 \nonumber\\
\alpha^{12}_{ab}&=& -{1\over 2}\bar{g}_8\sqrt{N}k
\langle\Delta\hat\lambda_a\rangle\delta_{0b} \nonumber\\
\alpha^{21}_{ab}&=&-{1\over 2}\bar{g}_8\sqrt{N}k
\langle\Delta\hat\lambda_b\rangle\delta_{0a}\, . 
\label{weakcon}
\ea
Here and in the following we use a short-hand notation for 
the octet and 27-plet operators as follows:
\be
k\cdot \Delta \equiv {4\over F^2} f^{ij}\Delta_{ij}~~~~f^{23}=f^{32}=1,\,\,
 0\, {\mbox{otherwise}}
\ee
for the octet case and 
\be
q\cdot\Delta\cdot\Delta\equiv  {4\over F^2} t^{ij,kl}\Delta_{ij}\Delta_{kl}
\ee
in the 27-plet case. The tensor $t^{ij,kl}$ has been defined in
Eq. (\ref{T27}) for $\Delta S=\pm 1$ and in Eq. (\ref{T272}) for
$\Delta S=2$.

The operators $N_\mu^-$ and $N_\mu^+$ admit the same decomposition as in
Eq. (\ref{DEC}). It gives
\ba
N^{-11}_{\mu\, ab}&=& {i\over 4}g_8k\langle\{\Delta, u_\mu\}[\hat\lambda_a,
\hat\lambda_b]\rangle
+{i\over 2} g_8k\langle
 \Delta (\hat\lambda_a u_\mu \hat\lambda_b-\hat\lambda_b
 u_\mu \hat\lambda_a )\rangle  \nonumber\\
&&\hspace{-0.8cm}
+ig_{27}q\langle\Delta[\hat\lambda_a,\hat\lambda_b]\rangle\langle\Delta
 u_\mu\rangle -{i\over 2}g_{27}q\langle[\Delta ,u_\mu ]\hat\lambda_a\rangle
\langle\Delta\hat\lambda_b\rangle
+{i\over 2}g_{27}q\langle[\Delta ,u_\mu ]\hat\lambda_b\rangle
\langle\Delta\hat\lambda_a\rangle \nonumber\\
&& \hspace{-0.8cm} + {i\over 2}\bar{g}_8 k
\langle\Delta[\hat\lambda_a,\hat\lambda_b]\rangle\langle u_\mu\rangle 
-{i\over 4} \bar{g}_8 \sqrt{N}k\langle\Delta [u_\mu ,\hat\lambda_a]\rangle
\delta_{0b}  
+{i\over 4}\bar{g}_8  \sqrt{N}k\langle\Delta [u_\mu ,\hat\lambda_b]\rangle
\delta_{0a} \nonumber\\ 
N^{-22}_{\mu\, ab}&=&0\nonumber\\ 
N^{-12}_{\mu\, ab}&=&{i\over 4}\bar{g}_8  \sqrt{N} k\langle\Delta [u_\mu
,\hat\lambda_a]\rangle \delta_{0b}  \nonumber\\ 
N^{-21}_{\mu\, ab}&=&-{i\over 4} \bar{g}_8 \sqrt{N}k \langle\Delta [u_\mu
,\hat\lambda_b]\rangle \delta_{0a}
\label{nm}
\ea
and
\ba
N^{+11}_{\mu\, ab}&=& -{i\over 4}g_8 k\langle [\Delta, u_\mu ]
\{\hat\lambda_a,\hat\lambda_b\}\rangle
 -{i\over 2} g_{27}q\langle[\Delta ,u_\mu ]\hat\lambda_a\rangle
\langle\Delta\hat\lambda_b\rangle   \nonumber\\
&& \hspace{-0.8cm} 
-{i\over 2}g_{27}q\langle[\Delta ,u_\mu ]\hat\lambda_b\rangle
\langle\Delta\hat\lambda_a\rangle
 -{i\over 4} \bar{g}_8 \sqrt{N}k\langle\Delta [u_\mu ,\hat\lambda_a]\rangle
\delta_{0b}  
-{i\over 4}\bar{g}_8 \sqrt{N} k\langle\Delta [u_\mu ,\hat\lambda_b]\rangle
\delta_{0a} \nonumber\\
N^{+22}_{\mu\, ab}&=&0\nonumber\\
N^{+12}_{\mu\, ab}&=&{i\over 4}\bar{g}_8\sqrt{N}k\langle\Delta [u_\mu
,\hat\lambda_a]\rangle \delta_{0b}  \nonumber\\ 
N^{+21}_{\mu\, ab}&=&{i\over 4}\bar{g}_8\sqrt{N}k\langle\Delta [u_\mu
,\hat\lambda_b]\rangle \delta_{0a}\, .
\label{np}
\ea
The weak operator $\hat{\omega}_{ab}$ and the strong operator
$\hat{\sigma}_{ab}$ can be decomposed in a more convenient manner as follows
\be
\hat{\Theta}_{ab} =\hat{\Theta}^{11}_{ab} {1+\tau_3\over 2}+
\hat{\Theta}^{22}_{ab} {1-\tau_3\over
  2}+\Delta\hat{\Theta}\delta_{a0}\delta_{b0} (1-\tau_1)\, ,
\ee
where the last term is a pure singlet contribution that mixes bosonic ghost and
physical singlet fields.
The components of the weak operator $\hat{\omega}_{ab}$  are 
\ba
\hat\omega^{11}_{ ab}&=& 
{1\over 8}g_8k\left (
\langle\{\Delta ,u^2\}\{\hat\lambda_a,\hat\lambda_b\}\rangle
-\langle\Delta ( \hat\lambda_a u^2 \hat\lambda_b+
\hat\lambda_b u^2 \hat\lambda_a)\rangle\right.\nonumber\\
&& \left. -\langle\{\Delta , u_\mu\}(\hat\lambda_a u^\mu \hat\lambda_b
+\hat\lambda_b u^\mu \lambda_a )\rangle 
 + \langle\Delta u_\mu\{\hat\lambda_a,\hat\lambda_b\} u^\mu 
\rangle\right ) \nonumber\\
&&+{1\over 8}g_8^\prime k\left ( 
\langle \{\Delta ,\chi_+\} \{\hat\lambda_a,\hat\lambda_b\}\rangle
-\langle [\Delta ,\chi_-] \{\hat\lambda_a,\hat\lambda_b\}\rangle \right )
\nonumber\\
&&+{1\over 4} g_{27} q\left ( \langle\Delta u_\mu\rangle\langle\{\Delta ,
u_\mu\}\{\hat\lambda_a,\hat\lambda_b\} \rangle 
-2\langle\Delta u_\mu\rangle\langle\Delta (\hat\lambda_a u^\mu
\hat\lambda_b +\hat\lambda_b u^\mu \lambda_a )\rangle \right.\nonumber\\
&& \left. + \langle[\Delta ,u_\mu ]\hat\lambda_a\rangle 
\langle[\Delta ,u^\mu ]\hat\lambda_b\rangle  \right ) \nonumber\\
&&-{1\over 4}\bar{g}_8\langle u_\mu \rangle k\langle\Delta (\hat\lambda_a
u^\mu \hat\lambda_b+\hat\lambda_b u^\mu \hat\lambda_a -{1\over 2}\{u^\mu ,\{
\hat\lambda_a,\hat\lambda_b\}\})\rangle  \nonumber\\
\hat\omega^{22}_{ ab}&=&0\nonumber\\
\Delta\hat\omega&=&
- N\left ( \tilde{v}_0 k\langle\Delta u^\mu\rangle
  \langle u_\mu \rangle  
+ \tilde{v}_8 k\langle\Delta u^2\rangle 
 + \tilde{v}_5k\langle\Delta\chi_+\rangle
+ \tilde{v}_{27} q\langle\Delta u_\mu\rangle\langle\Delta
u^\mu\rangle \right )\, ,\nonumber\\
&&
\label{omega}
\ea
while the strong operator $\hat\sigma_{ab}$ is given by 
\ba
\hat\sigma_{ab}^{11} &=& 
-{1\over 4}\langle [u_\mu ,\hat\lambda_a][u^\mu ,\hat\lambda_b]\rangle+
{1\over 4}\langle\{\hat\lambda_a,\hat\lambda_b\}\chi_+\rangle 
 \nonumber\\
\hat\sigma_{ab}^{22} &=& -{1\over
    4}\langle\{\hat\lambda_a,\hat\lambda_b\}4B_0{\cal M}\rangle 
 \nonumber\\
\Delta\hat\sigma &=&
 {N\over 3} (\alpha\Box +m_0^2) -N(v_1\langle u_\mu
  u^\mu\rangle +v_2\langle \hat\chi_+\rangle ) \, ,
\label{sigma}
\ea
where we have defined the subtracted operator  $\hat\chi_+ = \chi_+-4B_0
{\cal M}$, which gives 
$\langle\hat\chi_+\rangle =\langle\chi_+\rangle -2NM^2$, with $M$ the bare
meson mass, for degenerate quark masses.

In order to compute the ultraviolet divergences of the bosonic 
generating functional at
one loop given in Eq. (\ref{BOSGF}), the next step to
perform is to diagonalize the quadratic form $X^{\prime{}^T}D^\prime X^\prime$
by eliminating the mixing between
the singlet and non singlet components. We proceed in the same manner as it
was done in Ref. \cite{npb} for the purely strong sector. 
We decompose the quadratic form
in its non singlet, mixed and singlet sectors as follows
\be
X^{\prime T} D^\prime X^\prime = 
\overline{X}^T\, \overline{D}\, \overline{X} + X_0^T\, B^T\,
 \overline{X}
+\overline{X}^T\, B\, X_0 + X_0^T\, D_X\, X_0\, ,
\label{RESC}
\ee
where $\overline{D}$ acts on the non singlet graded vector $\overline{X}$, 
$B$ and $B^T$ are mixing operators between the non singlet $\overline{X}$ 
and singlet $X_0$ vectors and $D_X$ acts on the singlet graded vector $X_0$.
The mixing operator $B_{a0}$ and its transposed $B_{0a}$
are easily derived from the previous expressions, which give at order $G_F$
\be
B_{a0}= -{1\over 2}\{ N^-_{a0},d_s\}+{1\over
  2}[d_s,N^+_{a0}]+\hat{\sigma}_{a0} +\hat{\omega}_{a0}-{1\over 2}
 \left (\alpha_{al}\tau_3\hat{\sigma}_{l0} + \hat{\sigma}_{al}\tau_3 
\alpha_{l0} \right )-{1\over 2}[d_s,[d_s,\alpha_{a0} ]]\, .
\ee
By performing the translation 
\be
\overline{X} = \overline{X}^\prime - \overline{D}^{-1}B X_0\, ,
\label{SHIFT}
\ee
the quadratic form in Eq. (\ref{RESC}) is diagonalized to 
\be
 X^{\prime T} D^\prime X^\prime   = 
\overline{X}^{\prime{}^T } \overline{D}\, \overline{X}^\prime +X_0^T( D_X -
B^T\bar{D}^{-1} B)  X_0\, .
\ee
The first term on the r.h.s. is now a pure non singlet form, while in the
second term
the singlet operator $D_X$ is shifted via a non local differential term.
As already discussed in the case of strong interactions the non locality of
the singlet differential operator does not
prevent the derivation of the ultraviolet
 divergences of the one loop generating
functional in closed form, since they remain local. 
Denoting with $\overline{D}_X=D_X -B^T\bar{D}^{-1} B$ the shifted singlet 
operator, the complete 
quenched generating functional to one loop can now be formally 
written as follows
\be
e^{iZ^{\mbox{\tiny{q}}}_{\mbox{\tiny{1\,loop}}}}= {\cal N}{\det{D_\zeta}\over 
(\det{G\tau_3})^{1\over 2}(\det{\bar{D}})^{1\over 2}
(\det{\bar{D}_X})^{1\over 2}}\,
,
\ee
where in the derivation of the ultraviolet divergences we retain up to  order
 $G_F$ terms in the weak sector. The bosonic contribution to the generating
 functional has now been splitted in its singlet and non singlet part.
 
\subsection{ Integral over the non-singlet fields}

\label{BOSNSING}

The operator $\bar{D}_{ab}$ acts as follows ($a,b=1,\ldots N^2-1$):
\begin{eqnarray}
\bar{D}_{ab}\bar{X}^b &=& d_{s\mu } d^\mu_s\tau_3\bar{X}_a+
\left ({1\over 2}[d_s,N^+]_{ab}+\hat{\sigma}_{ab}+\hat{\omega}_{ab}
-{1\over 2}\left
  (\alpha_{al}\tau_3\hat{\sigma}_{lb} + \hat{\sigma}_{al}\tau_3 \alpha_{lb}
\right )\right.\nonumber\\
&&\left. -{1\over 2}[d_s,[d_s,\alpha_{ab} ]]\right )\bar{X}^b +O(G_F^2)
\nonumber\\
d_{s\mu} \bar{X}_a &=& \partial_\mu\bar{X}_a+\hat{\Gamma}_{\mu {ab}}\bar{X}^b 
\; \; ,
\label{D}
\end{eqnarray}
where we have written $\bar{D}_{ab}$ in terms of a rescaled covariant
derivative which now contains the weak operator $N^-_{ab}$;
the connection $\hat{\Gamma}_{\mu{ab}}$ is given by 
\begin{equation}
\hat{\Gamma}_{\mu{ab}}=-\langle \Gamma_\mu
[\hat\lambda_a,\hat\lambda_b ] \rangle \, {1+\tau_3\over 2}
 -{1\over 2} N^-_{\mu{ab}}\, ,
\end{equation}
which contains the usual strong interaction connection (first term),
where  $\Gamma_{\mu }= 1/2 ( [u^\dagger ,\partial_\mu u ]-i u^\dagger
r_\mu u -i u l_\mu u^\dagger )$, and a weak contribution (second term).
The projections of the graded operators $N_\mu^\pm ,\alpha , \hat\sigma
,\hat\omega$ onto the non-singlet subspace are easily derived from
Eqs. (\ref {weakcon}), (\ref{nm}), (\ref{np}), (\ref{omega}) and 
(\ref{sigma}).

By explicitly writing $\bar{D}_{ab}$ in terms of its projections onto the
physical and non physical subspaces one realizes that a mixing between the
physical $\xi$
and non physical $\tilde{\xi}$ fluctuations is still 
induced by the weak connection $\alpha_{ab}$. We get
\be
\bar{X}^T\bar{D}\bar{X}= \xi^T D_\xi\xi -\tilde\xi^T(\Box
-\hat{\sigma}^{22}) \tilde\xi
+\xi^T\left ({1\over 2}\alpha^{12}\hat\sigma^{22}\right )\tilde\xi
+\tilde\xi^T\left ({1\over 2}\hat\sigma^{22}\alpha^{21}\right )\xi\, , 
\ee
where the first term on the r.h.s. contains the differential operator 
$D_\xi$ which acts on the physical non-singlet field $\xi_a,\, a=1,\ldots
N^2-1$  and it is given by
\be
D_\xi= d^2 + {1\over 2}[d,N^{+11}]+\hat{\sigma}^{11}+\hat{\omega}^{11}
-{1\over 2}\{\alpha^{11},\hat{\sigma}^{11}\}
 -{1\over 2}[d,[d,\alpha^{11} ]] +O(G_F^2)\, .
\label{DXI}
\ee
The second term gives the action in the bosonic ghost non singlet sector and
the last two terms are mixing terms.
In the degenerate mass case that we are considering here 
${\cal M}=m_q{\bf{ 1}}$, the mixing terms are zero, 
since $\hat{\sigma}^{22}_{0a}=\hat{\sigma}^{22}_{a0}=0$,
so that the physical and non physical bosonic sectors
decouple  as it happens in the purely strong interaction
case. Also, since $ \hat{\sigma}^{22}_{ab}=M^2\delta_{ab}$, 
the $\tilde{\xi}$ action 
reduces to the free action which gives a trivial
multiplicative constant in the generating functional.

After these reductions the bosonic determinant in the non singlet sector
reduces to the determinant of $D_\xi$.
The ultraviolet divergent part of the integral over the $\xi$ fields can be
derived in closed form by regularizing the determinant in $d$
dimensions and using standard heat--kernel techniques. 
The result reads as follows:
\be
{i\over 2}\ln\det D_\xi = {i\over 2}\ln\det
D_\xi\vert_{\mbox{\tiny{strong}}} +  {i\over 2}\ln\det
D_\xi\vert_{(8)} + {i\over 2}\ln\det D_\xi\vert_{(27)}+
{i\over 2}\ln\det D_\xi\vert_{\Delta S=2}\, .
\ee
The first term was derived in Ref. \cite{npb} 
with the same notation adopted here 
 and we do not write it again. The remaining terms are the $\Delta S=\pm 1$
 weak contributions in the octet and 27--plet sector  and the $\Delta S=2$
 contribution, all in
 the presence of a singlet component of the meson field. We list in Appendix
 \ref{LIST} the definition of the ultraviolet divergent octet $W_i,\,
 \bar{W}_i$ and 27-plet $D_i,\, \bar{D}_i$ 
weak operators, where $\bar{W}_i$ and $\bar{D}_i$ are new operators induced by
the presence of the dynamical singlet field. 
The notation follows the one of Ref. \cite{EKW} for the octet case, with the
exception of the  $\bar{W}_i$ not introduced there and the eventual enlargement
of the basis for $N>3$. 
The calculation is done in $SU(N)$ with $N$ generic, so that we maintain
everywhere the explicit dependence upon the number of flavours.
We find:
\ba
&& {i\over 2}\ln\det D_\xi\vert_{(8)}= 
-\frac{1}{(4\pi)^2(d\!-\!4)} \int \! dx \, {N\over 16} \left\{ 
g_8^\prime \left ( W_5-W_9+W_{10}+W_{12}+W_{36}\right ) \right. \nonumber\\
&& +g_8\left ( {8\over 3}W_1
-{2\over 3}W_2 +W_5 +W_9  -W_{12} +{1\over 3}W_{14}+{2\over 3}W_{15}
 -{1\over 3}W_{16}-{1\over 6}W_{18} \right. \nonumber\\
&&\left. -{5\over 3}W_{19}+W_{20}+2W_{21}+2W_{22} +{2\over 3}W_{25}-W_{26}
+{1\over 6}W_{27} 
-W_{36}-{1\over 6}W_{37} -W_{38} \right ) \nonumber\\
&&\left. +{1\over 3}\bar{g}_8\left ( 2\bar{W}_1+2\bar{W}_2+3\bar{W}_3
-2\bar{W}_5-6\bar{W}_{8}-12\bar{W}_{9}+2\bar{W}_{11}+4\bar{W}_{15}
\right )\right\}
\nonumber\\
&&+{1\over 8}g_8^\prime \left (W_7+W_{11}-\overline{W}_3+\overline{W}_4\right
) + g_8\left ( 
{1\over 4}W_4+{1\over 8}W_6-{3\over 16}W_7+{1\over 16}W_8-{1\over 8}W_{11}
\right. \nonumber\\
&& -{5\over 24}\bar{W}_1+{1\over 6}\bar{W}_2+{1\over 16}\bar{W}_3-{1\over
  8}\bar{W}_4 -{1\over 24}\bar{W}_5
+{7\over 24}\bar{W}_{11}+{1\over 12}\bar{W}_{15}+{1\over 8}\bar{W}_{17}-
{1\over 8}\bar{W}_{13} \nonumber\\
&& \left. -{1\over 48}\bar{W}_8-{1\over 24}\bar{W}_9
-{1\over 8}\bar{W}_{12}+{1\over 8}\bar{W}_{16}-{1\over 12}\bar{W}_{14}
-{1\over 24}\bar{W}_{7}+{1\over 24}\bar{W}_{10}   \right )
\nonumber\\
&&+\bar{g}_8{1\over 8}\left ( -{1\over
    2}W_5-W_{10}+W_{38}+\bar{W}_{18}+2\bar{W}_{19} -\bar{W}_{20}-\bar{W}_{21}
-\bar{W}_{22}+\bar{W}_{23}\right )
\nonumber\\
&&+{1\over 8N}\left\{ g_8\left ( -W_5+2W_{10}+2W_{12}-4W_{21}-4W_{22}+2W_{36}
+2 W_{38}\right )\right.
\nonumber\\
&&\left. -2g_8^\prime\left ( W_{10}+W_{12}+ W_{36}\right ) 
+\bar{g}_8 W_{11} \right\} 
-{1\over 4N^2}(g_8- g_8^\prime ) W_{11} + O(G_F^2)\, .
\label{OCTETDET}
\ea
Notice that the octet determinant in the physical sector is the one in the
presence of the singlet field. The inclusion of the singlet field modifies the
standard ChPT determinant in three ways: the presence of 
the barred operators $\overline{W}_i$, the presence of new contributions to
the $W_i$
proportional to the singlet coupling $\bar{g}_8$
and an additional contribution to the $W_i$ counterterms coming from 
the anticommutator term
$\{\alpha^{11},\hat{\sigma}^{11}\}$  in Eq. (\ref{DXI}) when internal indices
run through zero.  The conversion from
one realization of ChPT to the other can be done using the following formula
\ba
&&\vspace{-0.6cm}{i\over 2}\ln\det D_\xi\vert_{(8)} =  {i\over 2}\ln\det
D_\xi\vert_{(8)}^{no\, singlet} -\frac{1}{(4\pi)^2(d\!-\!4)} \int \! dx \,
\left \{  g_8 \left [ {1\over 4N}W_{38}-{1\over 8N} W_5 \right.\right.
\nonumber\\
&&\left.\left.- {1\over 4N}W_{10} +{1\over 4N^2}W_{11}\right ]
+(g_8,g_8^\prime ,\overline{g}_8)\cdot
\overline{W}_i\, {\mbox{terms}}\, +\overline{g}_8\cdot W_i\, {\mbox{terms}}
\right\} +O(G_F^2)\, .
\ea
In actual computations of weak matrix elements with the singlet meson field
integrated out, the appropriate divergence of the  octet 
counterterms $W_i,\,
i=5,10,11,38$ has to be taken. This divergence is the one listed in Table
(\ref{TABCTR8}) in the CHPT column.
In the 27--plet sector we get
\ba
&& {i\over 2}\ln\det D_\xi\vert_{(27)}= -\frac{1}{(4\pi)^2(d\!-\!4)}
 \int \! dx ~ g_{27}\left\{ 
{N\over 12}\left ( D_1+D_2 -{1\over 2}D_4+{3\over 2}D_8-D_{13}
\right.\right.\nonumber\\
&&\left.+D_{14} -{1\over 2}D_{15}-D_{16}
+D_{17}+D_{18}+2D_{21}+2D_{22} \right )
+{1\over 12}D_1+{5\over 6}D_2-{1\over 8}D_3
\nonumber\\
&& -{7\over 24}D_4-{1\over 4}D_5 +{1\over 4}D_6
+{1\over 2}D_7+{3\over 8}D_8-{3\over 8}D_9+{1\over 4}D_{10}-{1\over 4}D_{11}
-{1\over 8}D_{12} 
\nonumber\\
&&-{1\over 12}D_{13} +{1\over 12}D_{14} -{1\over 24}D_{15} -{1\over 12}D_{16}
+ {7\over 12}D_{17}+  {7\over 12}D_{18} -{1\over 4}D_{19} -{1\over 4}D_{20}
 +{1\over 6}D_{21}\nonumber\\
&&\left. +{1\over 6}D_{22} +{1\over 4}D_{23}+{1\over 4}D_{24}
-{1\over 4}\bar{D}_1-{1\over 4}\bar{D}_2-{1\over 4}\bar{D}_3
 +{1\over 4N}D_{12}\right\} +{\cal O}(G_F^2)\, ,
\label{27DET}
\ea
where the appropriate tensor $t^{ij,kl}$ to $\Delta S=1$ or 2 interactions has to
be taken inside the $D_i$ and $\bar{D}_i$ counterterms defined in Appendix
\ref{LIST}. This time the presence of the singlet field modifies the standard
ChPT determinant only via the presence of the $\bar{D}_i$ counterterms.

\subsection{Integral over the singlet fields}

\label{SingInt}

After the shift of the $\overline{X}$ field as in Eq. (\ref{SHIFT})
the operator acting on $X_0$ can be written as follows:
\be
X_0^T \overline{D}_X X_0 = X_0^T \left[ D_X- B^T
  \bar{D}^{-1} B \right] X_0 \; \; ,
\ee
where
\ba
D_X &=& D_X^0+A \; \; ,\nonumber \\
D_X^0&=& \tau_3(\Box +M^2)+\frac{N}{3}(1-\tau_1)(\alpha \Box +m_0^2)
\; \; , \nonumber \\
A &=& A_s +A_w  \; \; , \nonumber \\
A_s &=& \frac{1}{4N}(1+\tau_3) \langle \hat\chi_+ \rangle
 -N(1-\tau_1)\left ( v_1 \langle u_\mu u^\mu \rangle +v_2 \langle
\hat\chi_+ \rangle \right ) +O(\Phi_0^2) \; \; , \nonumber \\
A_w  &=&   \frac{1}{4N}(1+\tau_3)(g_8^\prime -g_8)k\langle \Delta\chi_+ \rangle
-\left ({1+\tau_3\over 2}-{\tau_1\over 2}\right ) {\bar{g}_8\over 4}k
\langle \Delta\chi_+ \rangle    \nonumber \\ 
&&-N(1-\tau_1)\left (\tilde{v}_8 k\langle\Delta
u_\mu u^\mu\rangle +\tilde{v}_5 k\langle\Delta\chi_+\rangle
+\tilde{v}_{27}q\langle\Delta
u_\mu\rangle \langle\Delta u^\mu\rangle \right.   \nonumber \\ 
&& \left. +\tilde{v}_0 k
\langle\Delta u_\mu\rangle \langle u^\mu\rangle \right )+O(\Phi_0^2)
\; \; , \nonumber \\
&& \nonumber \\
B &=& B_s +B_w  \; \; , \nonumber \\
B_s^a &=& {1+\tau_3\over 2}\frac{1}{2\sqrt{2N}} \langle \lambda^a \chi_+ 
\rangle \; \; ,  
\nonumber \\
B_w^a &=&  {1+\tau_3\over 2}\frac{1}{4\sqrt{2N}}\left [g_8 k\left ( 
-i\langle [d^\mu_s,
 [\Delta ,u_\mu ]]\lambda^a\rangle -2\langle d_s^2\Delta\lambda^a\rangle 
-\langle \{\Delta ,\chi_+\}\lambda^a\rangle  \right.\right.\nonumber\\
&&\left. +{1\over 2}\langle [u^\mu , [\Delta ,u_\mu ]]\lambda^a\rangle
\right )
+g_8^\prime k\left ( \langle\{\Delta ,\chi_+\}\lambda^a\rangle
- \langle [\Delta ,\chi_-]\lambda^a\rangle \right )  \nonumber\\
&&\left. 
-{\bar{g}_8\over 2}k\langle\Delta\lambda^a\rangle\langle\hat{\chi}_+\rangle
-g_{27} q \langle\Delta\lambda^a\rangle\langle\Delta\chi_+\rangle
\right ] \nonumber\\ 
&&+\left ( {1+\tau_3\over 2}-{\tau_1+i\tau_2\over 2}\right
)\frac{N}{4\sqrt{2N}}\left [ -2g_8 k\langle\Delta\lambda^a\rangle\left (
{1\over 3}(\alpha\Box +m_0^2)-v_1\langle
u^2\rangle-v_2\langle\hat{\chi}_+\rangle\right ) \right. \nonumber\\ 
&& +\bar{g}_8 k\left ( {i\over 2} \langle\{ d^\mu_s,
 [\Delta ,u_\mu ]\}\lambda^a\rangle -{i\over 2}  \langle [d^\mu_s,
 [\Delta ,u_\mu ]]\lambda^a\rangle - \langle d_s^2\Delta\lambda^a\rangle
-M^2\langle\Delta\lambda^a\rangle \right. \nonumber\\  
&&\left.\left. 
+{1\over 4} \langle [u^\mu , [\Delta ,u_\mu ]]\lambda^a\rangle
-{1\over 4}\langle \{\Delta ,\chi_+\}\lambda^a\rangle  \right )
\right ]\, .
\label{DX}
\ea
The covariant derivative $d^\mu_s$ is the one defined in Eq. (\ref{COV}).
We separated the singlet operator $A$ and the mixing operator $B$ in their
strong interacting part (with subscript $s$) and their weak interacting part
up to order $G_F$. 
We recall that $\langle\hat\chi_+\rangle
=\langle\chi_+ \rangle -2NM^2$ in the degenerate mass case.
As also in the analysis of the purely strong interacting sector we 
cannot apply straightforwardly the heat--kernel
techniques, because the differential operator does not reduce to a
diagonal Klein--Gordon operator when the external fields are put to
zero. Therefore we just expand the logarithm of the differential
operator, and calculate only the ultraviolet divergent terms. The expansion
gives: 
\ba
{i\over 2}\mbox{Tr} \ln \left(\overline{D}_X/D_X^0 \right)  &=& 
{i\over 2}\mbox{Tr}\left[ {D_X^0}^{-1} (\overline{D}_X\!-\!D_X^0) \right]
\nonumber \\ 
&-&\frac{i}{4}
\mbox{Tr}\left[ {D_X^0}^{-1} (\overline{D}_X\!-\!D_X^0){D_X^0}^{-1}
  (\overline{D}_X\!-\!D_X^0) \right] + \ldots  \; ,
\label{logDX}
\ea
where the ellipsis contains ultraviolet finite terms only.
The inverse of the ``free'' operator $D_X^0$ is:
\be
{D_X^0}^{-1} = G_0 \left[ \tau_3 - (1+\tau_1) \frac{N}{3} ( \alpha
  \Box + m_0^2 ) 
  G_0 \right] \; \; ,
\label{DX0-1}
\ee
where
\be
( \Box +M^2 )_x G_0(x-y) = \delta(x-y) \; \; ,
\ee
and 
\be
\overline{D}_X-D_X^0 = A- B^T \bar{D}^{-1} B \;
\; . 
\label{ABX}
\ee
While the strong interacting part has been fully derived in Ref. \cite{npb},
the ultraviolet divergences of the singlet determinant
 at order $G_F$ in the expansion in powers of the weak coupling, 
are given by the following terms:
\ba
{i\over 2}\mbox{Tr} \ln \left(\overline{D}_X/D_X^0 \right)  &=& 
{i\over 2}\mbox{Tr}\left[ {D_X^0}^{-1}A_w\right ]
 -{i\over 2}\mbox{Tr}\left[{D_X^0}^{-1}A_s{D_X^0}^{-1}A_w\right ]
 \nonumber\\
&&\hspace{-2.5truecm} -{i\over 2}\mbox{Tr}\left[ {D_X^0}^{-1}B_s^T\bar{D}^{-1}B_w + 
 {D_X^0}^{-1}B_w^T\bar{D}^{-1}B_s \right ] \nonumber\\
&&\hspace{-2.5truecm} +{i\over 2}\mbox{Tr}\left[{D_X^0}^{-1}A_s {D_X^0}^{-1}\left
    (B_s^T\overline{D}^{-1}B_w +B_w^T\overline{D}^{-1}B_s\right )\right ]
+O(G_F^2)\, .
\label{UVEXP}
\ea
The inverse of the non singlet operator $\overline{D}$ is expanded around its
free part.
In the third term of Eq. (\ref{UVEXP}) we need its expansion 
 up to order $G_0^2$ and keeping only its strongly interacting part (since we
 stop at order $G_F$). This gives:
\be
\overline{D}^{-1}_{ab} = G_0\tau_3\delta_{ab}+G_0 \left ( M^2\delta_{ab}\tau_3
-\hat{\sigma}^{11}_{ab}{1+\tau_3\over 2}-\hat{\sigma}^{22}_{ab}
{1-\tau_3\over 2} \right ) G_0 +O(G_0^3)\, .
\ee
Note that the $O(G_0^2)$ term of $\overline{D}^{-1}_{ab}$ in the third term
of Eq. (\ref{UVEXP}) yields ultraviolet
divergences only when the term proportional to $\alpha\Box$ in $B_w$ is
considered. The same is true for the last term in Eq. (\ref{UVEXP}), where 
$\overline{D}^{-1}$ is only taken up to order $G_0$.
For the sake of clearness we list in Appendix \ref{SINGLET} the 
ultraviolet divergent contributions produced by each term of Eq. (\ref{UVEXP}).
The total ultraviolet divergent contribution given by the integral over the
singlet fields at order $G_F$ in the weak interaction sector is given by: 
\ba
&&\hspace{-0.6cm}
\frac{i}{2} \mbox{Tr} \ln \left( \overline{D}_X/D_X^0 \right) =\nonumber\\
&&\hspace{-0.6cm} -\frac{1}{(4\pi)^2 (d\!-\!4)} \int \! dx\,\,
\left\{ \frac{1}{6}m_0^2 (g_8^\prime -2g_8)k \langle\Delta\chi_+\rangle
+{\alpha^2\over 36} (g_8^\prime -g_8)k\langle\Delta\chi_+\rangle
\langle\hat{\chi}_+\rangle \right.\nonumber\\
&&\hspace{-0.6cm}+\frac{\alpha}{24}\left [ -2g_8^\prime (W_{10}+W_{12}+W_{36})
+g_8\left (4W_{10}
 +2W_{12}-4W_{21}-4W_{22}+2W_{36} \right )\right.
\nonumber\\
&&\hspace{-0.6cm}
\left. +2 g_{27}D_{12} + 2\bar{g}_8 k\langle\Delta\chi_+\rangle
 \langle\hat{\chi}_+\rangle\right ]
+{1\over 16}\bar{g}_8\left ( W_5-2W_{12}+4W_{21}+4W_{22}
 -2W_{36}-2W_{38}\right )\nonumber\\
&&\hspace{-0.6cm} -{1\over 2}\left (\tilde{v}_8 k\langle\Delta
u_\mu u^\mu\rangle +\tilde{v}_5 k\langle\Delta\chi_+\rangle
+\tilde{v}_{27}q\langle\Delta
u_\mu\rangle \langle\Delta u^\mu\rangle +\tilde{v}_0 k
\langle\Delta u_\mu\rangle \langle u^\mu\rangle \right )
\langle\hat{\chi}_+\rangle \nonumber\\
&&\hspace{-0.6cm} -{1\over 2} (g_8^\prime -2g_8)k \langle\Delta\chi_+\rangle
\left ( v_1\langle u_\mu u^\mu\rangle +v_2 \langle\hat{\chi}_+\rangle \right )
+{1\over 8N}\left [ 2g_8^\prime (W_{10}+W_{12}+W_{36})\right. \nonumber\\
&&\hspace{-0.6cm}
\left. +g_8\left ( W_5-2W_{10}-2W_{12}+4W_{21}+4W_{22}-2W_{36}
-2 W_{38}\right ) -\bar{g}_8W_{11} -2g_{27}D_{12}  \right ]
\nonumber\\
&&\hspace{-0.6cm}
\left. +{1\over 4N^2}(g_8-g_8^\prime )W_{11}\right\} + O(G_F^2)\, .
\label{SINGSING}
\ea
The most relevant behaviour concerns the appearance of the quenched chiral
logarithms, i.e. of the type $m_0^2\log m_\pi^2$, that are pure artefacts of
the quenched approximation. As it was discussed at length in Ref. \cite{npb}
 quenched chiral logarithms appearing in the strong sector can be formally
reabsorbed in a redefinition (in fact a true renormalization in dimensional
regularization) of the $B_0$ parameter. In the weak sector an analogous
 mechanism occurs. The quenched chiral logarithms which appear through the
 first term in
 Eq. (\ref{SINGSING})  can be formally reabsorbed into a redefinition of the
 weak mass term coupling $g_8^\prime$ of the leading order Lagrangian
 (\ref{WEAK}). To remove the $m_0^2$ divergence in 
 Eq. (\ref{SINGSING}) one has to add to the lowest order parameter 
 $g_8^\prime$ in the leading Lagrangian (\ref{WEAK}) a d-dependent part
 proportional to $m_0^2$ that has a pole at d=4:
\be
g_8^\prime\to g_8^\prime\left [1+{\mu^{d-4}\over 16\pi^2}{1\over d-4}
  {2m_0^2\over 3F^2}\left (1-2{g_8\over g_8^\prime}\right ) +\delta
  g_8^\prime (\mu )\right ]\, ,
\ee
so that the renormalized coupling can be defined as follows:
\be
g_{8R}^\prime =  g_8^\prime\left ( 1-{m_0^2\over 48\pi^2F^2}\left
    (1-2{g_8\over g_8^\prime}\right )\log {M^2\over \mu^2} +\delta
  g_8^\prime (\mu )\right )\, .
\label{QLOGS}
\ee   
The rescaling of the coupling $g_8^\prime$ together with the rescaling of
the parameter $B_0\to\bar{B}_0$ defined in Ref. \cite{npb} 
in the tree level contribution to
any weak observable can be used as a short-cut procedure to
unreveal the presence of quenched chiral logarithms, generated 
when the quenched approximation is implemented to one loop.

\subsection{The bosonic determinant: complete result}

\label{BOSCR}

The complete contribution to the bosonic part of the logarithm of the quenched 
generating functional to one loop is given by
\be
Z^b_{\mbox{\tiny{1\,loop}}}  = \frac{i}{2} \ln\det{\bar{D}} +
\frac{i}{2} \mbox{Tr} \ln \bar{D}_X 
+\frac{i}{2} \mbox{Tr} \ln (G\tau_3)\, ,
\ee
where the first term is the non singlet contribution which reduces to 
Eqs. (\ref{OCTETDET}) and (\ref{27DET}) in the degenerate mass case, 
the second term is the singlet part
given by Eq. (\ref{SINGSING}) and the last term comes from the 
Jacobian of the transformation induced by the weak connection. It gives:
\ba
&&\frac{i}{2} \mbox{Tr} \ln (G\tau_3)=\frac{i}{2} \mbox{Tr}\,
(\alpha_{ab}\tau_3)+O(G_F^2) =
\frac{i}{2}\int\, dx\, \alpha^{11}_{aa}(x) +O(G_F^2) = \nonumber\\
&&\frac{i}{2}\int\, dx\, 
\left ( Nk\langle\Delta\rangle +q\langle\Delta^2\rangle +\bar{g}_8
k\langle\Delta\rangle\right ) +O(G_F^2)\, ,
\ea
which is zero at order $G_F$ since the projection operator $\Delta_{ij}$ 
is traceless. 

It is easy to verify that the complete 
cancellation of the $1/N, 1/N^2$ terms is
provided by the sum of the non-singlet contributions (\ref{OCTETDET}), 
(\ref{27DET}) and the singlet contributions 
 (\ref{SINGSING}), a feature that was already pointed out in Ref. \cite{npb}
 for the strong sector. This cancellation is a consequence of the introduction
 of a dynamical singlet field, independently of whether the quenched
 approximation is made. Within the bosonic sector, quenching effects are of
 two types: 1) terms proportional to $m_0^2$ and $\alpha$ and 2) 
the substitution
 $\langle\chi_+\rangle\to \langle\hat{\chi}_+\rangle$. Terms of type 1) 
originate from the double-pole part of the super-$\eta^\prime$ two-point
 function, whereas the substitution of type 2) does eliminate the linear 
flavour number dependence that comes from $\langle{\cal M}\rangle$ in
 the degenerate mass case. By eliminating terms of
 type 1) and replacing back $\langle\chi_+\rangle$ the bosonic determinant 
$Z^b_{\mbox{\tiny{1\,loop}}}$ becomes the full 
ChPT determinant with a dynamical singlet field.

\subsection{The fermionic ghost sector}

\label{FER}

The differential operator $D_\zeta$ as defined in Eq. (\ref{S1}) 
is given by $(a,b=0,1,\ldots N^2-1)$
\ba
D_{\zeta{ab}}&=&d_\mu \bar{G}_{ab}d^\mu - {1\over  2}\{
\bar{N}^-_{ab},d\} + {1\over  2}[d,\bar{N}^+_{ab}]+\bar{\sigma}_{ab} 
+\bar{\omega}_{ab}\, ,
\ea
where $\bar{G}_{ab}=\delta_{ab}+\bar{\alpha}_{ab}$, $d_\mu$ is the covariant
derivative acting on the $\zeta$ field as defined in the purely strong
interaction case and the barred quantities are defined as follows
\ba
\bar{\alpha}_{ab}&=& {g_8\over
  2}k\langle\Delta\hat{\lambda}_a\hat{\lambda}_b\rangle \nonumber\\
\bar{N}^-_{\mu ab}&=&g_8 {i\over 4}k\langle\{\Delta
,u_\mu\}\hat{\lambda}_a\hat{\lambda}_b\rangle 
+g_{27} iq\langle\Delta\hat{\lambda}_a\hat{\lambda}_b\rangle \langle\Delta
u_\mu \rangle\nonumber\\
\bar{N}^+_{\mu ab}&=& -{i\over 4}g_8 k\langle [\Delta
,u_\mu ]\hat{\lambda}_a\hat{\lambda}_b\rangle \nonumber\\
\bar{\omega}_{ab}&=&  {g_8\over 8}k\langle (\{\Delta ,u^2\} +u_\mu\Delta u^\mu
) \hat{\lambda}_a\hat{\lambda}_b\rangle 
 +{g_{27}\over 4}q\langle\{\Delta
,u_\mu\}\hat{\lambda}_a\hat{\lambda}_b\rangle \langle\Delta
u^\mu \rangle \nonumber\\
&&+{g_8^\prime\over 8}k\left ( \langle \{\Delta ,\chi_+\} 
\hat{\lambda}_a\hat{\lambda}_b\rangle -\langle [\Delta ,\chi_-] 
\hat{\lambda}_a\hat{\lambda}_b\rangle \right )\nonumber\\
\bar{\sigma}_{ab}&=&{1\over 4}\langle (u_\mu u^\mu+\chi_++4B_0{\cal M})
 \hat{\lambda}^a\hat{\lambda}^b\rangle \, .
\ea
 After rescaling the coefficient of the double derivative term to unit, keeping
up to order $G_F$ terms and redefining the covariant derivative in the
presence of weak interactions (i.e. by reabsorbing the $\bar{N}_\mu^-$ term in
the covariant derivative $d_\mu$), the fermionic ghost determinant is now
given by the
product $\det D_\zeta = \det\bar{G}\cdot\det D^\prime_\zeta$, where the 
rescaled
operator $ D^\prime_\zeta$ acts on the fermionic ghost fields as follows 
\ba
D_{\zeta ab}^{\prime}\zeta^b&=&d_\mu d^\mu\zeta_a+\left ( {1\over
  2}[d,\bar{N}^+_{ab}]+\bar{\sigma}_{ab} +\bar{\omega}_{ab}-{1\over 2}\{\bar{
\alpha},\bar{\sigma}\}_{ab}-{1\over 2}[d,[d,\bar{\alpha}_{ab}]]\right )\zeta^b
\nonumber\\
d^\mu\zeta_a&=& \partial^\mu\zeta_a+\bar{\Gamma}^\mu_{ab}\zeta^b\, ,
\ea
with 
\be
\bar{\Gamma}^\mu_{ab}=-\langle\Gamma^\mu\hat{\lambda}_a\hat{\lambda}_b
\rangle -{1\over 2}\bar{N}^{-\mu}_{ab}\, .
\ee
The result for the ultraviolet divergences of the fermionic ghost determinant
at one loop reads as follows:
\ba
{i}\ln\det D_\zeta &=& {i}\ln\det
D^\prime_\zeta\vert_{\mbox{\tiny{strong}}} +  {i}\ln\det
D^\prime_\zeta\vert_{(8)} + {i}\ln\det D^\prime_\zeta\vert_{(27)}
+ {i}\ln\det D^\prime_\zeta\vert_{\Delta S=2}\nonumber\\
&& +i\mbox{Tr}\ln \bar{G}\, ,
\ea
where the first contribution from strong interactions has been given in
Ref. \cite{npb}, while the remaining contributions from the weak sector at
order $G_F$ are  
\ba
\hspace{-0.7cm} {i}\ln\det D^\prime_\zeta\vert_{(8)}&=& 
-\frac{1}{(4\pi)^2(d\!-\!4)} \int \! dx~ \left\{ 
D_\xi^{8}(N)+{N M^2\over 8}\left [
g_8 k\langle\Delta u^2\rangle \right.  \right.\nonumber\\
&&\left.\left. +2(g_8^\prime - g_8) k\langle\Delta\chi_+\rangle
  + 2\bar{g}_8 k \langle\Delta u_\mu\rangle \langle u^\mu \rangle 
\right ]\right\} +O(G_F^2)
\label{DETG8}
\ea
and for the 27-plet 
\ba
{i}\ln\det D^\prime_\zeta\vert_{(27)}&=& 
-\frac{1}{(4\pi)^2(d\!-\!4)} \int \! dx
~\left\{ D_\xi^{27}(N) 
 +{NM^2\over 2} g_{27}q  \langle\Delta u_\mu\rangle 
\langle\Delta u^\mu \rangle \right\} \nonumber\\
&& + O(G_F^2)\, .
\label{DETG27}
\ea
With the notation $D_\xi^{8}(N)$ and $D_\xi^{27}(N)$ we define the part of
the integrals over the physical $\xi$ field
in Eqs. (\ref{OCTETDET}) and (\ref{27DET}) that carries the
linear flavour number dependence.
The term ${i}\ln\det D^\prime_\zeta\vert_{\Delta S=2}$ is the same as in
Eq. (\ref{DETG27}) where the appropriate tensor $t^{ij,kl}$ is used. 
The contribution from the Jacobian term $i{\mbox{Tr}}\ln\bar{G}$ 
is again zero at order $G_F$, since it is given
by 
\ba
i\mbox{Tr}\ln \bar{G}&=&i\mbox{Tr}\,\bar{\alpha}_{ab}+O(G_F^2)=i\int\,
dx~\bar{\alpha}_{aa}+O(G_F^2)=  \nonumber\\
&&i\int\, dx~{g_8\over 2}Nk\langle\Delta\rangle +O(G_F^2)\, 
\ea
and $\Delta_{ij}$ is traceless. The way Eqs. (\ref{DETG8}) and (\ref{DETG27}) 
have been written shows explicitly that the integral over the fermionic ghost
fields generates a linear flavour number dependence only, as expected. 
As it was noted in Ref. \cite{npb}, the linear flavour number dependence of
the physical determinant is not fully explicit for degenerate quark masses.  
In fact, the
terms in addition to $D_\xi^{8}(N)$ and $D_\xi^{27}(N)$ in Eqs. (\ref{DETG8}),
(\ref{DETG27}) guarantee the
cancellation of those $N$ dependent  contributions generated in the physical
sector by the 
$\langle\chi_+\rangle$ in the degenerate mass case.

\renewcommand{\theequation}{\arabic{section}.\arabic{equation}}
\setcounter{equation}{0}
\section{Complete Result}

\label{CR}

In the analysis of the complete result for the 
ultraviolet divergences of the weak
generating functional to one loop, with and without the quenched
approximation,  we limit to the case of non-singlet Green's
functions. This corresponds to neglecting the barred chiral invariants
 $\overline{W}_i$
and $\overline{D}_i$ which we previously included in the partial
contributions.
They should be considered whenever singlet Green's functions are involved.
 We also remind the reader that the calculation of 
singlet Green's functions should also 
involve a set of contributions proportional to
$\overline\Phi_0$ ($\overline\phi_0$ in the unquenched case) 
 and coming from the expansion of the strong $V_i$ and weak 
$\tilde{V}_i$ potentials. For the sake of clearness we limit to the
non-singlet Green's functions that are the most 
relevant to phenomenological applications in the weak sector.

In the following formulas we give the complete ultraviolet divergent part of 
the hadronic weak generating functional (with the exclusion of singlet
operators) in the quenched approximation and with degenerate quark 
masses for $\Delta S=\pm
1$ interactions, octet and 27-plet, and for $\Delta S=2$ interactions.
They are as follows:
\ba
Z_{(8)}^q&=&
-\frac{1}{(4\pi)^2(d\!-\!4)} \int \! dx\, \left\{{1\over 6}m_0^2(g_8^\prime
  -2g_8) k \langle\Delta\chi_+\rangle
+{1\over 4}g_8 W_4 \right. \nonumber\\
&& +{1\over 8}W_6+\left [-{3\over 16}g_8 
\left ( 1-{16\over 3}v_1\right ) +{1\over
    8}g_8^\prime \left (1-4v_1\right ) \right ] W_7 
+{1\over 16}(g_8-8\tilde{v}_8)W_8  \nonumber\\
&&+\left ({\alpha\over 12}(g_8-g_8^\prime )-{1\over 8}\bar{g}_8\right ) 
\left ( W_{12}+W_{36}\right )
+\left ({\alpha\over 12}(2g_8-g_8^\prime )-{1\over 8}\bar{g}_8\right )W_{10} 
\nonumber\\
&&+ \left [ \left ({1\over 8}+{\alpha^2\over 36}\right )(g_8^\prime-g_8)
  -{1\over 2} v_2 (g_8^\prime -2g_8) -{1\over 2} \tilde{v}_5 +{\alpha\over
    12}\bar{g}_8 \right ] W_{11} \nonumber\\
&&\left.  + \left (-{\alpha\over 6}g_8+{1\over 4}\bar{g}_8\right ) 
(W_{21}+W_{22})  \right\} +O(G_F^2)\, ,
\ea
where counterterms $W_i$ are listed in Appendix \ref{LIST} and
the operators $W_8$ and $W_{11}$ have been replaced by the
quenched ones with the substitution
 $\langle\chi_+\rangle \to\langle\hat\chi_+\rangle $. As
explained in the case of strong interactions \cite{npb}, 
the redefinition
provides the subtraction of terms that become linearly dependent upon the
number of flavours in the degenerate quark mass limit. 
The 27-plet contribution is given by
\ba
Z_{(27)}^q&=&
-\frac{1}{(4\pi)^2(d\!-\!4)} \int \! dx\, g_{27}\left\{ 
{1\over 12}D_1+{5\over 6}D_2-{1\over 8}D_3-{7\over 24}D_4-{1\over 4}D_5
\right.\nonumber\\
&&+{1\over 4}D_6+{1\over 2}D_7+{3\over 8}D_8-{3\over 8}D_9
+{1\over 4}\left (
  1-2{\tilde{v}_{27}\over g_{27}}\right ) D_{10}
-{1\over 4}D_{11}
\nonumber\\
&&-{1\over 8}\left (1-{2\over 3}\alpha\right ) D_{12}
 -{1\over 12}D_{13} +{1\over 12}D_{14}-{1\over 24}D_{15}
-{1\over 12}D_{16}+ {7\over 12}D_{17}\nonumber\\
&&+  {7\over 12}D_{18} -{1\over 4}D_{19} -{1\over 4}D_{20}
+{1\over 6}D_{21}+{1\over 6}D_{22}\nonumber\\
&&\left. +{1\over 4}D_{23}+{1\over 4}D_{24} 
\right\}  +O(G_F^2)\, ,
\ea
where counterterms $D_i$ are listed in Appendix \ref{LIST} and
the operator $D_{10}$ has been replaced by the quenched one with 
 $\langle\chi_+\rangle \to\langle\hat\chi_+\rangle $.
The quenched generating functional for $\Delta S=2$ interactions has 
exactly the same
structure of $Z_{(27)}^q$, where the $\Delta S= 1$ 
tensor $t^{ij,kl}$ in the $D_i$ counterterms 
is replaced by the $\Delta S=2$ one.

\renewcommand{\theequation}{\arabic{section}.\arabic{equation}}
\setcounter{equation}{0}
\subsection{Analysis of divergent counterterms}

\label{DVCT}

In order to make clear within the present approach how the quenched
approximation modifies phenomenological predictions in the weak sector
we first investigate the main properties of the 
weak divergent counterterms in the full theory and their
r\^ole in mediating weak processes. We then compare the quenched value of each
divergent counterterm to the corresponding one in the full theory. Notice that
once the quenched counterpart of each divergent contribution has been
derived, one knows how the quenched approximation affects the coefficient of
chiral logarithms for any weak observable.

In the full theory
we define the divergent contribution to the $\Delta S=\pm 1$
 counterterm Lagrangian at order $p^4$ with the following sums
\be
{\cal L}^{4}_{(8)}= g_8\sum_{\{i\}} w_i W_i \, ,
\ee
\be
{\cal L}^{4}_{(27)}= g_{27}\sum_{i=1}^{24} d_i D_i \, ,
\ee
where the set of values $\{i\}$ in the octet Lagrangian 
runs over the divergent set of counterterms
given in Appendix \ref{LIST}. They are 25 in total. The 27-plet tensor
$t^{ij,kl}$ inside the $D_i$ invariants is the one defined in Eq. (\ref{T27}).
The counterterm Lagrangian in the $\Delta S=2$ case is 
\be
{\cal L}^{4}_{\Delta S=2}= g_{27}^{{\tiny{\Delta S=2}}}
\sum_{i=1}^{24} d_i D_i \, ,
\ee
 where now $t^{23,23}= t^{32,32}=1$ and 0 otherwise. 
Since the 27-plet operators which
 induce $\Delta S=1$ and $\Delta S=2$ transitions are components of the same
 irreducible tensor under $SU(3)_L\times SU(3)_R$, the 
 coefficients $d_i$ in both Lagrangians have to be the same.
 The renormalized value of the coefficients $w_i$ and $d_i$  
can be defined in the conventional way and using dimensional regularization:
\ba
w_i&=& \left (\upsilon_i + {g_8^\prime\over g_8}\upsilon_i^\prime
+{\bar{g}_8\over g_8}\bar{\upsilon}_i\right )\, \lambda +w_i^r(\mu )\equiv 
\nu_i\,\lambda + w_i^r(\mu )\nonumber\\
d_i&=& \delta_i\,\lambda + d_i^r(\mu )\, ,
\label{CTRDEF}
\ea
while we recall the analogous definition for the coefficients of the 
strong counterterms \cite{gl85}
\be
L_i = \Gamma_i\,\lambda +L_i^r(\mu )\, .
\label{CTRSDEF}
\ee 
$\lambda$ contains the divergence at $d=4$
\be
\lambda = \frac{\mu^{d-4}}{16
  \pi^2}\left[\frac{1}{d-4}-\frac{1}{2}\left( \ln 4\pi
    +\Gamma^\prime(1)+1\right) \right] \; \; ,
\label{LAMBDA}
\ee
 $\mu$ is the renormalization scale, while the coefficients
$\upsilon_i,\,\upsilon_i^\prime ,
 \,\bar{\upsilon}_i,\, \delta_i$ and their
quenched counterpart $\upsilon_i^q,\,\upsilon_i^{\prime q} ,
 \,\bar{\upsilon}_i^q,\, \delta_i^q$
are given in Tables (\ref{TABCTR8}) and  (\ref{TABCTR27}). The coefficients
 $\Gamma_i$ of the strong counterterms and their quenched counterpart can be
 found in Ref. \cite{npb}. For the following phenomenological analysis 
it is also useful to introduce the scale independent constants
\ba
\bar{w}_i&=&{32\pi^2\over \nu_i} w_i^r(\mu ) -\ln{M^2\over\mu^2}\nonumber\\
\bar{d}_i&=&{32\pi^2\over \delta_i} d_i^r(\mu ) -\ln{M^2\over\mu^2}\nonumber\\
\bar{L}_i&=&{32\pi^2\over \Gamma_i} L_i^r(\mu ) -\ln{M^2\over\mu^2}
\label{BARREDCTR}
\ea
and their quenched counterparts $\bar{w}_i^q,\,\bar{d}_i^q,\, \bar{L}_i^q$.
 They carry the chiral logarithms, e.g. $\bar{w}_i =
-\ln M^2 +\,\ldots$, where $M^2$ is the squared bare meson mass, and the
analogous definition for the other constants.

The quenched approximation largely reduces the ultraviolet
 divergent contribution to
the octet sector at one loop. The only octet operators whose divergence is not
modified by quenching are $W_i,\,  i=4,6$. Within the class of divergences
proportional to $g_8$, 
the octet operators $W_i,\, i=10,12,21,22,36$ acquire a divergence proportional
to $\alpha$ coming from the quenched anomalous singlet sector and $W_{11}$
gets a contribution proportional to $\alpha^2$. $W_7,\, W_8$ and $W_{11}$ get
a contribution from the strong and weak potentials $v_i,\, \tilde{v}_i$.
Notice that our list of octet counterterms in the full theory 
differs from the one in Ref. \cite{EKW}
for the presence of $W_{38}$, which has to be kept as a linearly independent
operator for a flavour group $SU(N)$  with generic $N$.
For $N=3$ flavours this operator can be
 eliminated with the use of Cayley-Hamilton relations, that give $W_{38} =
 -W_5+W_6+1/2 W_7+W_8$.

The octet operators $W_i$ contribute to several different decay processes of
the $K$ meson. $W_1,\,\ldots W_4$ contribute to $K\to3\pi$ decays, 
$W_5,\,\ldots W_{12}$ and $W_{38}$ to $K\to 2\pi ,\, 3\pi$ decays, 
$W_{14},\,\ldots W_{18}$ to radiative $K$ decays and finally 
$W_{19},\,\ldots W_{27}$ only to processes which involve an external $W$ gauge
boson (in addition to the non leptonic weak transition). $W_{36}$ and
$W_{37}$ are contact terms which are only needed for renormalization purposes.

The r\^ole of the contributions induced by the weak mass term 
(i.e. the terms proportional
to $g_8^\prime$) has been extensively discussed in Ref. \cite{BPP} for the most
general case of off-shell Green's functions, while in Ref. \cite{EKW} it is shown
how their contribution can be reabsorbed in a rescaling of some of the 
counterterms proportional to $g_8$ for on-shell matrix elements.
The operators $W_i,\, i=5,7,9,10,11,12,36$ get a contribution proportional to
$g_8^\prime$ in the full theory. In the quenched approximation the
contribution to $W_5$ and $W_9$
becomes zero, while $W_{10},\, W_{12},\, W_{36}$ get a contribution 
proportional
to $\alpha$, $W_{11}$ to $\alpha^2$ and $W_7,\, W_{11}$ to $v_i,\,
\tilde{v}_i$.

We also included the contributions coming from the 
extra term proportional to $\bar{g}_8$ of the leading
$O(p^2)$ weak Lagrangian in the presence of the singlet field. 
This term induces contributions to the octet  
operators $W_i,\, i=10,11,12,21,22,36$, in addition to all the singlet induced 
barred operators $\overline{W}_i$. After quenching, only the divergence of
$W_{11}$ is modified (with the exclusion of the $\bar{W}_i$ operators), 
where a contribution proportional to $\alpha$ appears. 

In the 27-plet sector all the unquenched chiral invariants $D_i$ survive to
the quenched approximation.
The operators $D_i,\, i=1,2,4,8,13,14,15,16,17,18,21,$ $22$ loose their linear 
flavour number dependent contribution, $D_{12}$ looses the $1/N$ term due to the
presence of the dynamical singlet field, while 
getting a contribution proportional to $\alpha$. $D_{10}$ gets a contribution
from $\tilde{v}_{27}$.
Counterterms $D_1,\,\ldots D_7$ contribute to  $K\to3\pi$ decays,
$D_8,\,\ldots D_{12}$  to $K\to 2\pi ,\, 3\pi$ decays, $D_{13},\,\ldots
D_{16}$ to radiative $K$ decays and $D_{17},\,\ldots D_{24}$ to 
processes with an external $W$ gauge boson.

\begin{table}
\begin{center}
\begin{tabular}{|c|c|c|c|c|c|}
\hline
 &\multicolumn{2}{c|}{}&\multicolumn{2}{c|}{}&\\
$W_i$ &\multicolumn{2}{c|}{$\upsilon_i$}& \multicolumn{2}{c|}
{$\upsilon^\prime_i$}
 &$\bar{\upsilon}_i$  \\
 &\multicolumn{2}{c|}{CHPT~~~~~~qCHPT} &\multicolumn{2}{c|}{CHPT
~~~~~qCHPT} &   qCHPT \\
&\multicolumn{2}{c|}{\phantom{CHPT}~~~~~~~$\langle \chi_+ \rangle
  \to \langle \hat{\chi}_+ \rangle$}& 
\multicolumn{2}{c|}{\phantom{CHPT}~~~~~~$\langle \chi_+ \rangle
  \to \langle \hat{\chi}_+ \rangle$}& $\langle \chi_+ \rangle
  \to \langle \hat{\chi}_+ \rangle$\\
\hline
 &&&&&\\
1 & ${N\over 6}$ & 0 & 0 & 0  & 0\\
2 & $-{N\over 24}$ & 0 & 0 & 0 & 0 \\  
4 & ${1\over 4}$& ${1\over 4}$ & 0 & 0  & 0 \\
5 & ${N\over 16}$ 
& $ 0$ & ${N\over 16}$ & 0 & 0  \\
6 & ${1\over 8}$ & ${1\over 8}$ & $0$ &  $0$ & $0$  \\
7 & $-{3\over 16}$ & $-{3\over 16}\left (1-{16\over 3}v_1\right )$ 
&${1\over 8}$   &${1\over 8}(1-4v_1)$  & 0 \\
8 & ${1\over 16}$ & ${1\over 16}\left (1-8{\tilde{v}_8\over g_8}\right )$  
& $0$ & 0 & 0 \\
9 & ${N\over 16}$ & 0 & $-{N\over 16} $ & 0  &  0 \\
10 & ${1\over 2N}$& ${\alpha\over 6}$ 
& ${N\over 16}- {1\over 4N}$ &$-{\alpha\over 12}$  &
 $-{1\over 8}$  \\
11 & $-{1\over 8}-{1\over 2N^2}$
& $-{1\over 8}-{\alpha^2\over 36}+v_2$ & ${1\over 8}+{1\over 4N^2}$ 
& ${1\over 8}+{\alpha^2\over 36}-{v_2\over 2}-{\tilde{v}_5\over 2g_8^\prime}$
 & ${\alpha\over 12}$ \\
12 & $-{N\over 16}+{1\over 4N} $ &${\alpha\over 12}$  
& ${N\over 16}-{1\over 4N} $ &$-{\alpha\over 12}$  & $-{1\over 8}$  \\
14 &  ${N\over 48}$ & 0 & $0$ & 0  & 0  \\
15 &  ${N\over 24}$ & 0 & $0$ & 0 & 0 \\
16 &  $-{N\over 48}$ & 0 & $0$ & 0 & 0 \\
18 &  $-{N\over 96}$ & 0 & $0$ & 0  & 0 \\
19 &  $-{5\over 48}N$ & 0 & $0$ & 0  &0  \\
20 &  ${N\over 16}$ & 0 & $0$ & 0   & 0 \\
21 &  ${N\over 8}-{1\over 2N}$ &$-{\alpha\over 6}$  & $0$ & 0 
  & ${1\over 4}$   \\
22 &  ${N\over 8}-{1\over 2N}$ &$-{\alpha\over 6}$  & $0$ & $0$ 
  & $ {1\over 4}$ \\
25 &  ${N\over 24}$ & 0 & $0$ &0  & 0 \\
26 &  $-{N\over 16}$ &0  & $0$ &0 & 0 \\
27 &  ${N\over 96}$ & 0 & $0$ &0  &0  \\
36 & $-{N\over 16}+{1\over 4N} $ &  ${\alpha\over 12}$
 & ${N\over 16}-{1\over 4N} $ & $-{\alpha\over 12}$  &  $-{1\over 8}$ \\
37 &  $-{N\over 96}$ &0  & $0$ & 0 & 0 \\
38 & $-{N\over 16}$ &  $0$ & $0 $ &0  & 0 \\
&&&&&\\
\hline
\end{tabular}
\end{center}
\protect\caption{List of the octet counterterms $W_i$ at order $p^4$ for 
a generic number of flavours $N$.
For each counterterm we list the coefficients of the divergent part
 $\upsilon_i$, $\upsilon_i^\prime$ and $\bar{\upsilon}_i$ as defined in 
Eq. (\protect\ref{CTRDEF}) in the unquenched case with no singlet field
(denoted as CHPT) and in the quenched case (qCHPT). 
The unquenched contribution from $\bar{\upsilon}_i$ is always zero in standard
CHPT where no dynamical singlet field is present.
As we have indicated in the table, chiral invariants
  containing $\langle \chi_+ \rangle$ have to be changed with $\langle
  \chi_+ \rangle \to \langle \hat{\chi}_+ \rangle$ in the quenched case.}   
\label{TABCTR8}
\end{table}

\begin{table}
\begin{center}
\begin{tabular}{|c|c|c|}
\hline
 &\multicolumn{2}{c|}{}\\
$D_i$ &\multicolumn{2}{c|}{$\delta_i$}\\
 &\multicolumn{2}{c|}{CHPT~~~qCHPT} \\
&\multicolumn{2}{c|}{\phantom{CHPT${}_s$}~~~~$\langle \chi_+ \rangle
  \to \langle \hat{\chi}_+ \rangle$}\\
\hline
 &&\\
1 & ${N\over 12}+{1\over 12} $ &${1\over 12}$  \\
2 & ${N\over 12}+{5\over 6}$ &${5\over 6}$ \\  
3 & $-{1\over 8}$ &  $-{1\over 8}$ \\
4 & $-{N\over 24}-{7\over 24}$ &$-{7\over 24}$ \\   
5 & $-{1\over 4}$ &  $-{1\over 4}$ \\  
6 & ${1\over 4}$ &  ${1\over 4}$ \\  
7 & ${1\over 2}$ &  ${1\over 2}$ \\  
8 & ${N\over 8}+{3\over 8} $ &$ {3\over 8} $  \\
9 & $-{3\over 8}$ &  $-{3\over 8}$ \\  
10 & ${1\over 4}$ & ${1\over 4}\left (1-2{\tilde{v}_{27}\over g_{27}}\right )$
 \\  
11 & $-{1\over 4}$ &  $-{1\over 4}$ \\  
12 & $-{1\over 8}+{1\over 4N}$ &  $-{1\over 8} \left (1-{2\over 3}\alpha\right
 )$ \\  
13 & $-{N\over 12}-{1\over 12} $ &$-{1\over 12}$  \\
14 & ${N\over 12}+{1\over 12} $ &${1\over 12}$  \\
15 & $-{N\over 24}-{1\over 24}  $ &  $-{1\over 24}$ \\  
16 & $-{N\over 12}-{1\over 12} $ &$-{1\over 12}$  \\
17 & ${N\over 12}+{7\over 12} $ &${7\over 12}$  \\
18 & ${N\over 12}+{7\over 12} $ &${7\over 12}$  \\
19 & $-{1\over 4}$ &  $-{1\over 4}$ \\  
20 & $-{1\over 4}$ &  $-{1\over 4}$ \\  
21 & ${N\over 6}+{1\over 6} $ &${1\over 6}$  \\
22 & ${N\over 6}+{1\over 6} $ &${1\over 6}$  \\
23 & ${1\over 4}$ &  ${1\over 4}$ \\  
24 & ${1\over 4}$ &  ${1\over 4}$ \\  
&&\\
\hline
\end{tabular}
\end{center}
\protect\caption{List of the 27-plet divergent counterterms 
$D_i$ at order $p^4$ 
for $N$ generic. Notation as in Table (\protect\ref{TABCTR8}).}
\label{TABCTR27}
\end{table}

\renewcommand{\theequation}{\arabic{section}.\arabic{equation}}
\setcounter{equation}{0}
\section{Applications to weak observables}

\label{phen}

In this section we use the results obtained through the calculation of
the full and quenched weak generating functional to one loop to estimate in
a few cases how the quenched approximation modifies the  
contribution coming from chiral logarithms to nonleptonic weak matrix
 elements.
We focus on $B_K$ in the $\Delta S=2$ sector and $K\to \pi\pi$ matrix elements
 in the $\Delta S =1$ sector. The $B_K$ parameter has
been extensively investigated in the literature both in the unquenched case
(see e.g. Refs. \cite{GOLKPP,BKKPP,BPBK,Sharpe}) 
and in the quenched approximation 
(see e.g. Refs. \cite{Sharpe,GOLKPP,BKQUENCHED} and Refs. 
\cite{BKLAT,BKLAT2} for recent lattice determinations). 
Recent attempts to compute unquenched (really partially quenched) $B_K$ on 
the lattice can be found in Refs. \cite{K1,K2,K3}.
The same interest has been devoted to the formulation of 
 analytic approaches to $K\to \pi\pi$ decays in the full theory, see
e.g. \cite{BURAS,BPP,KPPFULL,A1,A2,PASCHOS}. 
An attempt to fix the
order $p^4$ counterterms in $K\to \pi\pi$ decays from the
 off-shell $K\pi ,\, K\eta$ weak transitions
 has been done in Ref. \cite{BPP}.
In the quenched approximation, a first 
analysis of the $\Delta I=3/2$ decay $K^+\to\pi^+\pi^0$ can be found 
in Ref. \cite{GOLKPP}, where finite volume effects on the lattice are also
investigated. The most recent numerical determination of $K^+\to\pi^+\pi^0$
amplitude on the lattice and in the quenched approximation is reported in
Ref. \cite{AOKI}. 
In what follows we shall concentrate on the
comparison of chiral logarithms that contribute to the various weak 
quantities in
the full and in the quenched theory. Although some of the results are 
well known, as in the case of $B_K$,
 the structure of the counterterms and their flavour number
dependence will be clear in this context. 

\renewcommand{\theequation}{\arabic{section}.\arabic{equation}}
\setcounter{equation}{0}
\subsection{The $B_K$ parameter}

\label{BKpar}

The $B_K$ parameter is defined in terms of $K_0\bar{K}_0$ amplitudes as follows
\be
{\cal M}_K = \langle K_0\vert O_K(x)\vert \bar{K}_0\rangle = B_K(\mu )\, {\cal
  M}_{vac}\, .
\ee
 $O_K(x)$ is the effective four-quark operator $O_K(x) = (\bar{s}(x)\gamma_\mu
(1-\gamma_5)d(x))  (\bar{s}(x)\gamma^\mu (1-\gamma_5)d(x))$, 
where summation over colours is understood within brackets, 
 and ${\cal M}_{vac}$ is the result of the vacuum saturation approximation
\be
{\cal M}_{vac} = {8\over 3} \langle K_0\vert \bar{s}\gamma_\mu
(1-\gamma_5)d\vert  0\rangle \langle 0\vert \bar{s}\gamma^\mu (1-\gamma_5)d
\vert \bar{K}_0\rangle \, .
\ee
The scale dependence of $B_K(\mu )$ is due to the fact that the effective
four-quark operator $O_K(x)$ has an anomalous dimension. The scale independent
quantity is the physical matrix element $\langle K_0\vert {\cal
  H}_{\tiny{ \Delta S=2}}\vert \bar{K}_0\rangle$, where the effective
Hamiltonian ${\cal H}_{\tiny{ \Delta S=2}}$ can be written as 
${\cal H}_{\tiny{ \Delta S=2}} = -C^{\tiny{\Delta S=2}}
    \hat{O}_K(x)$, with the constant $C^{\tiny{ \Delta S=2}}$ defined
      in Eq. (\ref{CDS2}), while the matrix element of the four-quark operator 
$\hat{O}_K(x)$ is now scale invariant and defines the renormalization group
invariant $\hat{B}_K = B_K(\mu )\alpha_S(\mu )^{a_+}$ \cite{ANOM}. 
The realization of $\hat{O}_K(x)$ in terms of low energy degrees of freedom 
is contained in the effective Lagrangian of Eq. (\ref{DS2}) that gives 
$\hat{O}_K(x) = G_{27} F^4
\langle\Delta_{32}u_\mu\rangle\langle\Delta_{32}u_\mu\rangle $.
At order $p^2$ in the chiral expansion one obtains
\be
{\cal M}_K=4 G_{27}F^2m^2\,
 ,~~~~~~~~{\cal M}_{vac}={16\over 3}F^2m^2\, ,
\ee
where $F$ and $m^2$ are the bare kaon decay constant and squared mass and 
\be
\hat{B}_K = {3\over 4} G_{27}\, ,
\ee
with $G_{27}=1$ in the large-$N_c$ limit\footnote{At order $p^2$ in ChPT 
the same parameter $G_{27}$ relates $B_K$ to the $\Delta I=3/2$
  amplitude $K^+\to\pi^+\pi^0$ \cite{BKKPP}. One loop corrections due to
  $SU(3)$ breaking were derived in \cite{BSW}.}. 
The same result is valid in the quenched approximation where the bare
parameter $G_{27}$, the squared meson mass $m^2$ and the meson
 decay constant $F$ are replaced by the quenched ones.
At order $p^4$ in the chiral expansion
 the vacuum saturation amplitude gets renormalized so that 
\be
 {\cal M}_{vac}={16\over 3}F_K^2m_K^2\, .
\ee
The full amplitude ${\cal M}_K$ 
receives contributions from the strong and weak sectors.
A detailed analysis at order $p^4$ in ChPT can be found in Ref. \cite{BPBK}. 
Within the
$1/N_c$ expansion one can distinguish the factorizable (i.e. the non zero ones
in the large $N_c$ limit) and the non-factorizable diagrams at order $p^4$. 
The factorizable
contributions provide the renormalization of masses and decay
constants. The non-factorizable contributions are the only ones relevant to
$B_K$. We write $B_K$ as the sum $B_K =
B_K\vert_{ctr}+B_K\vert_{logs}$, where the first term contains the analytic
contributions coming from the effective Lagrangian up to order $p^4$ and the
second term contains the non-analytic corrections coming from the one-loop
diagrams. The generating functional as it was derived in the previous sections
gives in one step the divergent part of the analytic contribution to any
hadronic weak matrix element; in other words it gives the coefficients
of the chiral logarithms that contribute at one loop 
to any hadronic weak matrix element,
both in the full theory and in the quenched theory. We limit the analysis to
 $B_K\vert_{ctr}$ with the purpose of illustrating 
how the quenched approximation modifies the coefficients of the chiral 
logarithms in the degenerate mass case. At order $p^4$ 
in the full theory $B_K\vert_{ctr}$ reads as follows
\be
B_K\vert_{ctr} = {3\over 4} G_{27}\left\{1+{m_K^2\over F^2}\left (A+B+C\right
  )\right\}\, ,
\label{BKU}
\ee
with 
\ba
A&=& -16\,
 d_{12}\left ( 1-{m_\pi^2\over m_K^2}\right )^2 
\nonumber\\
B&=& 8(d_{10}-2 L_4){2B_0\langle{\cal M}\rangle\over m_K^2} + 16(d_8-L_5)
 \nonumber\\ 
C&=& -16\, d_f\, ,
\ea
where $d_8,\, d_{10},\, d_{12}$ are the weak $\Delta S=2$ counterterms 
defined in
Eq. (\ref{CTRDEF}) and $L_4,\, L_5$ are the strong counterterms as defined in
Eq. (\ref{CTRSDEF}) (see also
Ref. \cite{npb} for their definition in the full and quenched
case). The contribution from $C$ is due to one finite counterterm
$q\langle\Delta\chi_-\rangle\langle\Delta\chi_-\rangle$ with coefficient $d_f$.

In the
degenerate quark mass case (i.e. ${\cal M}=m_q{\bf 1}$) $A$ is zero, while the
first coefficient in $B$
develops an extra linear flavour number dependence. Notice
that the combinations $d_{10}-2 L_4$ and  $d_8-L_5$ in $B$ 
 come from the residual 
non-factorizable part of the total contribution from counterterms 
$D_{10}$ and $D_8$, while
their factorizable part goes into the renormalization of $F_K$ in the full
${\cal M}_K$ matrix element. In addition,
the flavour number dependence of $D_8$ and $D_{10}$ is exclusively contained
in their factorizable part.
It is now a simple exercise to show that in the degenerate quark mass case
 the chiral logarithms contributing  to 
$B_K$ at one loop are the same in the full theory and in the quenched theory 
(i.e. no flavour number dependence is produced in the full theory for
degenerate quark masses). 
In the quenched degenerate mass case the operator $D_{10}$ does not 
contribute anymore to $B_K$ so that at order $p^4$ one gets
\be
B_K^q\vert_{ctr} =  {3\over 4} G_{27}^q\left\{1+{M_K^2\over  F^{q2}}
\left (A^q+B^q+C^q\right )\right\}\, ,
\label{BKQ}
\ee
with 
\ba
A^q&=& 0
\nonumber\\
B^q&=& 16(d^q_8-L^q_5)\nonumber\\ 
C^q&=& -16\, d_f^q\, 
\ea
and the weak coupling, mass and decay constant are the quenched ones.
Notice also that $L_5^q$ in the quenched case becomes finite.

The renormalized $B_K\vert_{ctr}$ can be obtained from  Eq. (\ref{BKU})
through the substitution $L_i\to{\Gamma_i\over
32\pi^2}\bar{L}_i$ and $d_i\to{\delta_i\over 32\pi^2}\bar{d}_i$ (with the
exception of $d_f$), where the barred constants are the scale independent
constants defined in Eq. (\ref{BARREDCTR}).

The renormalized $B_K^q\vert_{ctr}$ can analogously be written in terms of 
$\bar{L}_i^q$ and $\bar{d}_i^q$ through the same substitution 
(with the exception of $L_5^q$) in Eq. (\ref{BKQ}).
 The scale independent constants carry the chiral logarithm,
e.g. $\bar{d}_i = -\ln M^2 +\,\ldots\,$.
By inserting in the renormalized expressions of $B_K\vert_{ctr}$ and
$B_K^q\vert_{ctr}$ the values of the 
coefficients $\delta_i$ and $\delta_i^q$ as given in Table
(\ref{TABCTR27}) and using $\Gamma_4 = 1/8,\, \Gamma_5=N/8$ for
the coefficients of the divergences in $L_4$ and $L_5$  
 we get the contribution to $B_K$ coming from the chiral
logarithms in the full and quenched theory
\ba
B_K\vert_{ctr}&=&  {3\over 4} G_{27}\left\{1-6\,{ m_K^2\over 32\pi^2
    F^2}\ln{M^2\over\mu^2} +\ldots\,\right\}
\nonumber\\ 
B_K^q\vert_{ctr}&=&  {3\over 4} G_{27}^q\left\{1-6\,{ M_K^2\over 32\pi^2
    F^{q2}}\ln{M^2\over\mu^2} +\ldots\,\right\}\, ,
\label{BKctr}
\ea
where the renormalization scale dependence cancels in the sum.
Eq. (\ref{BKctr}) proves the flavour independence of 
the next-to-leading contribution to
$B_K$ in the full theory and the consequent equality of the coefficients of
chiral logarithms in the full and quenched case for degenerate quark
masses\footnote{No contribution from the anomalous singlet sector is present
  in the quenched degenerate mass case.}.

\renewcommand{\theequation}{\arabic{section}.\arabic{equation}}
\setcounter{equation}{0}
\subsection{$K\to \pi\pi$ matrix elements}

\label{kppMAT}

The analysis of $K\to \pi\pi$ matrix elements in the
full theory, using the effective weak chiral Lagrangian at order $p^4$ has
been done in Ref. \cite{BPP}. Using those formulas and the quenched counterpart
of the divergent counterterms derived in the previous sections, we can produce
a few quantitative estimates of quenching effects on the coefficients of  
chiral logarithms that contribute to $K\to \pi\pi$ amplitudes. We work in the
infinite volume limit for illustrative purpose, while we defer to future work
the analysis of aspects more closely related to an actual 
lattice simulation.
We consider $K\to \pi\pi$ matrix elements with $\Delta I=1/2$ and 
$\Delta I=3/2$ in the full theory at one loop 
and derive the modifications induced by quenching in
the coefficients of chiral logarithms for degenerate light quark masses 
$m_u=m_d\equiv \hat{m}, \hat{m}= m_s$ (i.e. $m_K =m_\pi$).
 The amplitudes in the full theory are first computed in the case where no 
dynamical
singlet component is present and for non degenerate quark masses. 
Next, we consider the two limits $m_K =m_\pi$ and $m_\pi =0$. In the
 first case, $m_K =m_\pi$, the explicit flavour number dependence of the 
full amplitudes is shown. Once the flavour number dependence of the 
full amplitudes is known in the degenerate mass case, it is immediate to 
derive the corresponding quenched expression for degenerate masses.
Each coefficient of the chiral logarithms in the full amplitudes is replaced
by its quenched value according to Tables \ref{TABCTR8} and \ref{TABCTR27}. 
This  in practice amounts to eliminate the flavour number dependence of the
 full amplitudes and to add new
contributions (proportional to $m_0^2,\, \alpha ,\, v_i,\, \tilde{v}_i$) 
coming from the anomalous singlet sector.

We decompose the 
$K\to\pi\pi$ matrix elements into
definite isospin invariant amplitudes as follows
\ba
A \left[ K_S \to \pi^0 \pi^0 \right] \equiv
\sqrt{\frac{2}{3}}A_0 - \frac{2}{\sqrt 3} A_2 \, ; \nonumber \\
A \left[ K_S \to \pi^+ \pi^-\right] \equiv
\sqrt{\frac{2}{3}}A_0 + \frac{1}{\sqrt 3} A_2 \, ; \nonumber \\
A \left[ K^+ \to \pi^+ \pi^0\right] \equiv
\frac{\sqrt 3}{2} \, A_2 \, ,
\ea
where $K_S \simeq K^0_1 + \varepsilon \, K^0_2$, 
$K^0_{1(2)} \equiv (K^0-(+)\overline{K^0})/\sqrt 2$, 
CP $K^0_{1(2)} = +(-)K^0_{1(2)}$ and we set 
 $\varepsilon =0$ since $CP$ violation is small.
The isospin 1/2 amplitude can be written as follows
\ba
A_0 \equiv -i a_0 \, e^{i \delta_0} \, 
\ea
and the analogous for the isospin 3/2 amplitude 
\ba
A_2 \equiv - i a_2 \, e^{i \delta_2} \, ,
\ea 
where $\delta_{0,2}$ are final state interaction phases.
At order $p^2$ we get 
\ba
\Im m A_0&\equiv& \Im m (A_0^8+ A_0^{27})= - \left[ g_8 + \frac{1}{9} g_{27}
 \right]
\, {\sqrt{6}\over F^3}\, (m_K^2-m_\pi^2) \,  , \nonumber \\
\Im m A_2&=& -  \, g_{27} 
\, \frac{10 \sqrt 3}{9 F^3} \, (m_K^2-m_\pi^2) \,  , 
\label{TREEkpp}
\ea
where we use $F = 93$ MeV and 
\ba
\delta_0 = \delta_2 = 0 \, . 
\ea
In the quenched approximation each parameter in Eq. (\ref{TREEkpp})
(i.e. masses, weak couplings and decay constant) has to be replaced by its
quenched counterpart.
At order $p^4$ in the chiral expansion we define the full generic amplitude 
$\Im m A_i$ as the sum $\Im m A_i\vert_{ctr} + \Im m A_i\vert_{logs}$, where
the first term contains the analytic contributions coming from the effective
Lagrangian up to order $p^4$ and the second term contains the non-analytic
corrections coming from the one-loop diagrams. Both contributions have been
derived for the physical amplitudes in Ref. \cite{BPP}.
In the following we generalize the analytic part of the amplitudes to a 
generic number of flavours; one can then obtain its quenched
approximation by replacing each divergence in the full
amplitude by its quenched value and setting to zero any residual flavour
number dependence (i.e. 
the one  coming from $\langle{\cal M}\rangle$ contributions at
degenerate quark masses).
This procedure is the most immediate to obtain the
 coefficients of chiral logarithms in the full and quenched theory
and in the degenerate mass case. 

\begin{figure}
\begin{center}
\leavevmode\epsfxsize=12cm\epsfbox{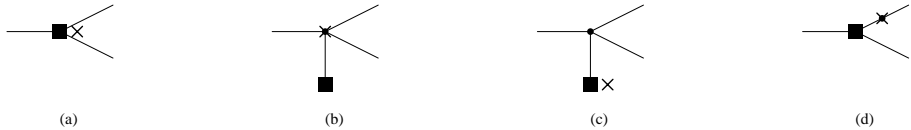}
\end{center}
\caption{Tree level diagrams that contribute to $K\to\pi\pi$ matrix elements
  at order $p^4$. The crossed box is the insertion of a weak counterterm
  (octet $w_i$ or 27-plet $d_i$). The crossed dot is the insertion of a
  strong counterterm $L_i$. The box is an order $p^2$ 
 weak vertex ($g_8,\, g_8^\prime ,\,  g_{27}$). The dot is an order $p^2$
  strong vertex. 
Diagram (d) is a wave function renormalization diagram for each external leg.
\label{figctr}}
\end{figure}

For the octet physical amplitude in the full theory, with $\hat{m}\neq
m_s$, we get
\ba
{\Im}m A_0^8\vert_{ctr}&=& -g_8{\sqrt{6}\over
  F_KF_\pi^2}(m_K^2-m_\pi^2 )\left \{ 1+{2\over F^2}\left [ 
4 m_\pi^2\left (2w_5+4w_7-2w_{10} \right. \right. \right.
\nonumber\\
&&\hspace{-1.5truecm}\left.\left.\left. -4w_{11}-2w_{12} +w_{38}\right )
 +4 m_K^2\left (w_5-2w_7+w_9\right ) 
+ 8 B_0\langle{\cal M}\rangle  w_8\right ]\right\} \nonumber\\
&&\hspace{-1.5truecm}-g_8^\prime  {8\sqrt{6}\over F_KF_\pi^2}
{(m_K^2-m_\pi^2 )\over F^2}\left [
m_\pi^2\left (-4L_4-L_5+8L_6+4L_8\right )+2m_K^2 L_4\right ]\, ,
\label{A8}
\ea
where the meson decays constants $F_K$ and $F_\pi$ in the leading order $p^2$
amplitude are the renormalized ones; their expression at order $p^4$ is given
in Refs. \cite{gl85,BPP}; their analytic contribution at order $p^4$ 
for a generic $N$ gives the following result
\be
F_K F_\pi^2\vert_{ctr}
 = F^3 \left [1+{24 B_0\langle{\cal M}\rangle\over F^2} L_4+
{4\over F^2}(m_K^2+2m_\pi^2) L_5 \right ]\, .
\label {FREN}
\ee
In Figure (\ref{figctr}) we list the tree level diagrams  which give
contribution to all $K\to\pi\pi$ amplitudes up to order $p^4$ in the 
chiral expansion. The first two lines of Eq. (\ref{A8}) come from diagrams
$(a)$ and $(c)$ of Figure (\ref{figctr}). Diagram $(c)$
 gives contribution only to
$w_{11}$. The third line comes from
diagram $(b)$, while diagram $(d)$ for each external line  gives 
the renormalization of $F_KF_\pi^2$ in the leading order $p^2$ amplitude. 
 Meson masses $m_K$ and $m_\pi$ are
always the renormalized physical ones because they come from factors $p_K^2$,
$p_\pi^2$, where we take the external momenta on shell. 
The expression in Eq. (\ref{A8}) is valid for a generic number of 
flavours $N$. In this respect it
differs from the one in Ref. \cite{BPP}.
 Also, our choice of the
 octet basis slightly differs from the one adopted in there, while it
 coincides with the one chosen in Ref. \cite{EKW} when counterterm 
$W_{38}$ is eliminated in the three flavours case. 
All counterterms in Eq. (\ref{A8}) are divergent.
In the degenerate quark mass limit the counterterm $W_8$ induces a linear
flavour number dependence due to the term 
$2B_0\langle{\cal M}\rangle\to  Nm_\pi^2$.
Besides, $W_{11}$ yields a linear flavour number dependence in the degenerate
mass case, being proportional to $\langle\chi_+\rangle$. However this
contribution drops out
in the sum of diagrams $(a)$ and $(c)$ depicted in Fig. (\ref{figctr}).
The strong counterterms with coefficients $L_4$ and $L_5$ 
again produce a linear
flavour number dependence in the wave function renormalization diagrams of the
type shown 
in $(d)$ of Figure (\ref{figctr}), 
which is fully reabsorbed in the renormalization of the product $F_K F_\pi^2$
of meson decay constants.

For instructive purpose we derive the flavour number dependence of the
coefficients of chiral logarithms that enter the octet amplitude 
at order $p^4$ and in the degenerate 
quark mass case. To this aim we first obtain 
the fully renormalized analytic contribution at order $p^4$ 
 by performing the
substitution $w_i\to{\nu_i\over 32\pi^2} \bar{w}_i$ and $L_i\to{\Gamma_i\over
  32\pi^2}\bar{L}_i$ in Eq. (\ref{A8}), where $\bar{L}_i,\,\bar{w}_i$ are the
scale independent constants defined in Eq. (\ref{BARREDCTR}).
Then we use the fact that $\bar{w}_i = -\ln m_K^2
+\ldots$, $\bar{L}_i = -\ln m_K^2+\ldots$ to derive the coefficients of chiral
logarithms. Their contribution is given
by the following ratio  
\be
{{\Im}m A_0^8\vert_{ctr}\over {\Im}m A_0^8\vert_{tree}}\vert_{\hat{m}=m_s}
= 1- \left ( {1\over 2}-{6\over N}+{8\over N^2} \right )
 {m_K^2\over 16\pi^2F^2}   \ln{m_K^2\over \mu^2}+\,\ldots\,\, ,
\label{A8div}
\ee
where we expressed everything in terms of $m_K^2$ and the $\mu$ dependence
cancels in the complete amplitude.
${\Im}m A_0^8\vert_{tree}$ is the order $p^2$ bare amplitude of
Eq. (\ref{TREEkpp}), also obtained from Eq. (\ref{A8}) when
the product $F_KF_\pi^2$ has been converted
to the unrenormalized $F^3$ through the relation (\ref{FREN}); the divergent
contribution to (\ref{FREN}) is obtained using
$\Gamma_4 = 1/8$, $\Gamma_5 = N/8$ for the coefficients of the divergence in
$L_4$ and $L_5$.
Notice that the chiral logarithm coming from the strong counterterms part
 $g_8^\prime\cdot L_i$ in Eq. (\ref{A8}) 
exactly cancels the one proportional to $g_8^\prime$ coming from the 
weak counterterms $w_i$'s. For $N=3$ the coefficient of the chiral logarithm
in Eq. (\ref{A8div}) is $(11/18)$, which becomes $-(35/9)$ when
$F^3\to F_KF_\pi^2$ is replaced in the tree level amplitude\footnote{ Notice
  that the coefficient of the chiral logarithm in Eq. (\ref{A8div}) becomes 
$-(5/4)$ in the non degenerate case for $m_\pi =0$. This can be derived from 
Eqs. (\ref{A8}) and (\ref{FREN}) by setting $m_\pi =0$.}.

The analytic contributions to the 
27-plet physical amplitudes up to order $p^4$ are as follows:
\ba
{\Im}m A_0^{27}\vert_{ctr}&&= -g_{27}{\sqrt{6}\over 9
  F_KF_\pi^2}(m_K^2-m_\pi^2 )\left \{ 1+{1\over F^2}\left [ 
8 m_\pi^2\left (2d_8- 2d_f+6d_9\right.\right.\right.
\nonumber\\
&&\left.\left.\left. -6d_{12}\right )
 +4 m_K^2\left (d_8-9d_9+d_{11}\right ) 
+ 16B_0\langle{\cal M}\rangle d_{10}\right ]\right\} 
\label{A27} \\
{\Im}m A_2\vert_{ctr}&&= -g_{27}{10\sqrt{3}\over 9
  F_KF_\pi^2}(m_K^2-m_\pi^2 )\left \{ 1+{1\over F^2}\left [ 
 16 m_\pi^2 (d_8- d_f) \right.\right.
\nonumber\\
&&\left.\left.  +4 m_K^2\left (d_8+d_{11}\right ) 
+ 16B_0\langle{\cal M}\rangle d_{10}\right ]\right\} \, .
\label{A2}
\ea
In the 27-plet case there is one finite counterterm with coefficient $d_f$
which gives contribution. This counterterm is $q\langle\Delta\chi_-\rangle
 \langle\Delta\chi_-\rangle$ and it does not produce any flavour number
 dependence. 
As in the octet case, counterterm $D_{10}$ yields a linear
 flavour number dependence in the degenerate mass case.
Again, the renormalized analytic contribution is obtained by
 performing the substitution $d_i\to {\delta_i\over 32\pi^2}\bar{d}_i$ (with
 the exception of the finite term $d_f$) in Eqs. (\ref{A27}) and (\ref{A2}).
The flavour number dependence of the coefficients of chiral logarithms 
in the 27-plet case is now given by the following ratios
\be
{{\Im}m A_0^{27}\vert_{ctr}\over {\Im}m
  A_0^{27}\vert_{tree}}\vert_{\hat{m}=m_s}  = 
1 + \left ( -{3\over 4}N -4+{6\over N} \right ) 
{m_K^2\over 16\pi^2F^2} \ln{m_K^2\over \mu^2} +\ldots\, ,
\label{A27div}
\ee
which gives $-(17/4)$ for the coefficient of the chiral logarithm with
$N=3$ (it goes to $-(35/4)$ when $F^3\to F_KF_\pi^2$) and
\be
{{\Im}m A_2\vert_{ctr}\over {\Im}m A_2\vert_{tree}}\vert_{\hat{m}=m_s}  = 
1 + \left ( -{3\over 4}N-{13\over 4} \right )
{m_K^2\over 16\pi^2F^2} \ln{m_K^2\over \mu^2} +\ldots\, ,
\label{A2div}
\ee
which gives $ -(11/2)$ for the coefficient of the chiral logarithm with
$N=3$ (it goes to $-10$ when $F^3\to F_KF_\pi^2$). The non degenerate
case with $m_\pi =0$ moves the coefficients of chiral logarithms to
$-(15/2)$ and $-(3/4)$ in Eq. (\ref{A27div}) and (\ref{A2div})
respectively.

The quenched version of the analytic contribution to the ${\Im}m A_i$ 
amplitudes at order $p^4$ and in the degenerate mass case can be easily derived
by replacing the quenched values for the coefficients of the divergences
$\nu_i,\,\delta_i,\,\Gamma_i$,  
 as given in Ref. \cite{npb} and Tables (\ref{TABCTR8}), 
(\ref{TABCTR27}) and dropping the residual flavour number dependence due to
the $\langle{\cal M}\rangle$ for degenerate masses.
Again, we define the limit of equal masses of the counterterm over tree
amplitude ratio and we define as $M$ the degenerate meson mass in the quenched
amplitudes. The quenched octet ratio is given by:
 \ba
{{\Im}m A_0^8\vert^Q_{ctr}\over {\Im}m A_0^8\vert^Q_{tree}}\vert_{\hat{m}=m_s} 
 &=&  1+{2M^2\over F^{2}}\left
   (12w_5^q+8w_7^q+4w_9^q-8w_{10}^q-16w_{11}^q-8w_{12}^q +4w_{38}^q\right )
\nonumber\\
&&\hspace{-1.5truecm}+{g_{8}^{\prime}\over g_8} {8M^2\over F^{2}}
\left (-2L_4^q-L_5^q+8L_6^q+4L_8^q\right )\, ,
\label{A8Q}
\ea
where the weak order $p^2$ couplings and decay constant are the 
quenched ones. Notice that in the quenched case the product of meson decay
constants in the leading order amplitude is only renormalized at one loop by a
finite counterterm \cite{npb}. It
is important to note that the weak mass term contribution from $g_8^\prime$
never appears in the  leading order amplitude, but it gives a
residual contribution of the type $g_8^{\prime}\cdot L_i^q$ to (\ref{A8Q}). 
As we saw in Section (\ref{SingInt}) 
the quenched chiral logarithms generated in the
weak interaction sector can be formally reabsorbed into a redefinition of the
weak coupling $g_8^\prime\to g_{8R}^\prime $. In addition, 
the anomalous behaviour of the
singlet sector in the quenched approximation yields a modification in the
power counting due to the presence of a new dimensionful expansion parameter
$m_0^2$, the singlet squared mass. For this reason, quenched ChPT 
becomes a double expansion in powers of the standard chiral parameter 
$\epsilon \equiv M^2/16\pi^2F^2$ and
the singlet parameter $\delta \equiv  m_0^2/16N_c\pi^2F^2$ (which induces
a $1/N_c$ expansion).
 Eq. (\ref{A8Q}) shows that in the degenerate mass case
quenched chiral logarithms cannot appear at one
 loop, while the one obtained from the rescaling of $g_8^\prime\to 
g_{8R}^\prime$ in Eq. (\ref{A8Q}) is already a two loop effect and of 
order $\epsilon\cdot \delta$ in the combined chiral and $1/N_c$
expansion\footnote{Notice however
 that the order $p^2$ amplitude always contains
 the fully renormalized meson masses.} . 

The 27-plet quenched ratios are as follows:
 \ba
{{\Im}m A_0^{27}\vert^Q_{ctr}\over {\Im}m
 A_0^{27}\vert^Q_{tree}}\vert_{\hat{m}=m_s}  &=&  1+{M^2\over F^2}\left
   (20d_8^q-16d_f^q+12d_9^q+4d_{11}^q-48d_{12}^q\right ) \\
\label{A27Q}
{{\Im}m A_2\vert^Q_{ctr}\over {\Im}m
 A_2\vert^Q_{tree}}\vert_{\hat{m}=m_s}  &=&  1+{M^2\over F^2}\left
   (20d_8^q-16d_f^q+4d_{11}^q\right )\, .
\label{A2Q}
\ea
As it happens in the full theory only one finite counterterm
$q\langle\Delta\chi_-\rangle\langle\Delta\chi_-\rangle$ with coefficient
$d_f^q$ contributes to the 27-plet amplitudes.
 
The contribution from the chiral logarithms to the quenched amplitudes
can be derived with the same
procedure adopted in the full case, i.e.
 through the derivation of the renormalized
amplitudes in terms of the scale independent constants $\bar{w}_i^q,\,
\bar{d}_i^q,\,\bar{L}_i^q$ and using the quenched
values of the coefficients $\nu_i,\, \delta_i$  as given in Tables
(\ref{TABCTR8}) and (\ref{TABCTR27}). For the coefficients of the divergence
in the strong $\bar{L}_i^q$ we use $\Gamma_4^q=1/8-v_1/2 ,\,
\Gamma_6^q= 1/16-v_2/2+\alpha^2/72,\, \Gamma_8^q=-\alpha/12 $,
as derived in Ref. \cite{npb}. The result reads as follows:
 \ba
{{\Im}m A_0^8\vert^Q_{ctr}\over {\Im}m A_0^8\vert^Q_{tree}}\vert_{\hat{m}=m_s} 
 &=&  1- \epsilon\,\left [
 {1\over 2}+8v_1-16v_2+{4\over
     9}\alpha^2-2\alpha   +8{g_8^\prime\over g_8}\left (
  -v_2+{\tilde{v}_5\over g_8^\prime}\right )      \right.\nonumber\\
&&\left. 
+2{\bar{g}_8\over g_8}\left ( 1-{2\over 3}\alpha\right )\right ] \ln{M^2\over
\mu^2}  +\ldots\,\, , \\
\label{A8Qdiv}
{{\Im}m A_0^{27}\vert^Q_{ctr}\over {\Im}m
 A_0^{27}\vert^Q_{tree}}\vert_{\hat{m}=m_s}  &=&  1- 4
\left (1-{1\over 2}\alpha\right ) \epsilon\, \ln{M^2\over \mu^2}+\ldots\,\, ,\\
\label{A27Qdiv}
{{\Im}m A_2\vert^Q_{ctr}\over {\Im}m
 A_2\vert^Q_{tree}}\vert_{\hat{m}=m_s}  &=&  1- {13\over 4} \epsilon\,
 \ln{M^2\over \mu^2}+\ldots\,\, ,
\label{A2Qdiv}
\ea
where we defined $\epsilon = M^2/16\pi^2 F^2$.
While in the $\Delta I=3/2$ amplitude $A_2$ no trace of the singlet anomalous
sector is present through chiral logarithms proportional to the singlet 
parameters $\alpha ,\, v_i$ and $\tilde{v}_i$, this is not the case for
the $\Delta I=1/2$ amplitudes also in the degenerate quark mass limit.
\begin{figure}
\begin{center}
\leavevmode\epsfxsize=12cm\epsfbox{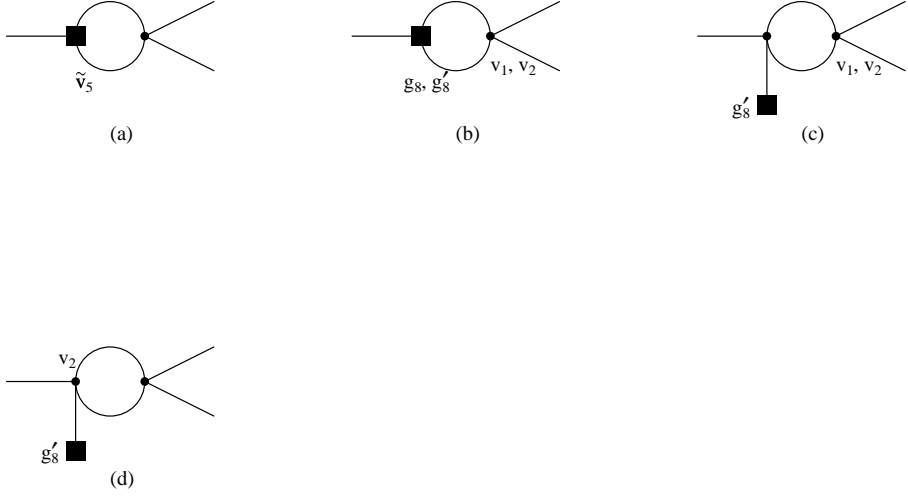}
\end{center}
\caption{One loop contributions to $K\to\pi\pi$ matrix elements with a single
  insertion of a weak $\tilde{v}_i$ or a strong $v_i$ vertex in the quenched
  approximation. A singlet field is always running in the loop. Notation is as
  in Figure (\ref{figctr}).
\label{figvi}}
\end{figure}
In Figure (\ref{figvi}) we clarify the origin of the one loop contributions
to Eq. (\ref{A8Qdiv}) that are 
proportional to the strong singlet potentials $v_1,\, v_2$ and the weak
singlet potential $\tilde{v}_5$. No contribution proportional to $\tilde{v}_8$
nor $\tilde{v}_{27}$ is allowed in this case. 
The total contribution to Eq. (\ref{A8Qdiv}) and 
proportional to  $g_8^\prime\cdot v_1$ 
 is zero due to the exact
cancellation of loops (b) and (c) in Figure (\ref{figvi}).
For $v_2$ the extra one loop contribution shown in (d) of Figure (\ref{figvi})
gives the net divergence in $A_0^8$ in the quenched case.
Contributions proportional to
$\alpha$ and $\alpha^2$ are the usual ones coming from one or 
two insertions of the double pole term of the singlet propagator in a one loop
diagram (no more than one insertion per singlet line).

In Table (\ref{QFULL}) we compare the coefficients of chiral logarithms for
the three amplitudes in the full theory and in the quenched approximation.
For the full amplitudes we consider two extreme mass configurations in the one
loop corrections: a) degenerate masses, i.e. $m_K=m_\pi$ and b) $m_\pi =0$.
The numerical analysis of the physical non degenerate mass case together with
the comparison with {\em unphysical} choices of the matrix elements on the
lattice will be given in Ref. \cite{E2}.
\begin{table}
\begin{center}
\begin{tabular}{|c|c|c|c|}
\hline
 & Full ($m_K=m_\pi$) &  Full ($m_\pi =0$) & Quenched \\
\hline
 &&&\\
  $A_0^8$   & ${11\over 18}$ & $-{5\over 4}$  & $-\left ({1\over 2}+{4\over
 9}\alpha^2 -2\alpha\right )+v_i,\tilde{v}_i,\bar{g}_8,g_8^\prime$ \\
 &&&\\
  $A_0^{27}$   & $-{17\over 4}$ & $-{15\over 2}$  & $-4
\left (1-{\alpha\over 2}\right )$ \\
 &&&\\
  $A_2$   & $-{11\over 2}$ & $-{3\over 4}$  &$-{13\over 4}$ \\
 &&&\\
\hline
\end{tabular}
\end{center}
\protect\caption{ The coefficient of $m^2/(16\pi^2F^2)\ln (m^2/\mu^2)$
  (with $m^2=m_K^2$ (Full), $m^2=M^2$ (Quenched)) for the
  full amplitude with $m_K=m_\pi$ (second column), $m_\pi =0$ (third column)
  and for the quenched amplitude (fourth column).}
\label{QFULL}
\end{table}
Notice that the 
coefficient of the chiral logarithm in the quenched octet amplitude
cannot be determined because of its dependence 
 upon unknown parameters of the singlet sector. 

The coefficients of chiral logarithms in the full 27-plet amplitudes ${\Im}m
 A_0^{27}$ and  ${\Im}m A_2$ are quite large in the degenerate mass
  limit. In addition, by the comparison of second and third column in 
Table (\ref{QFULL})
one can conclude that all the coefficients in the full amplitudes are
  extremely sensitive to the variation of masses. Going from the degenerate
  mass case to the limit $m_\pi =0$ the coefficient of 
the $\Delta I=3/2$ amplitude decreases in absolute value 
by a large amount (from -5.5 to -0.75, $\sim 86\%$), while the one of 
the $\Delta I=1/2$ amplitude increases by $\sim 76\%$ (from -4.25 to -7.5)
in the 27-plet case. The coefficient of the octet case changes sign going from
 0.6 to -1.25.  
Going from the case $m_K = m_\pi$ to the case $m_\pi =0$ we notice a drastic 
increase of the corrections to the $\Delta I=1/2$ amplitudes, in contrast to 
a sizable decrease of the corrections to the  
 $\Delta I=3/2$ amplitude. The values of the coefficients at  $m_\pi =0$
 clearly produce an enhancement of the $\Delta I=1/2$ amplitude.
In Ref. \cite{E2} a more
  complete analysis of the physical case will be given.

Setting $\alpha\simeq 0$ and disregarding for now the unknown contributions to
the octet amplitude in the quenched case we find the following pattern going 
from the full degenerate mass case  to the quenched one.
Quenching reduces from 0.6 to -0.5  the coefficient of
the chiral logarithm in the octet amplitude $A_0^8$. Notice however that the 
quenched result is very sensitive to the numerical value of the parameter
 $\alpha$. At present, 
no precise measurement of $\alpha$ is available on the lattice.

The coefficient in $A^{27}_0$ is only reduced by about $6\%$ in absolute value 
(from -4.25 to -4). Again, a value of $\alpha$ different from zero affects 
the result. For example,  $\alpha = 0.6$ would give a reduction of about 
 $34\%$ (from -4.25 to -2.8). Comparing instead with the  $m_\pi =0$ full 
case, quenching, with  $\alpha = 0$, reduces the coefficient in  $A^{27}_0$ 
by about $47\%$ (from -7.5 to -4).
In the case of $A_2$, comparing with the full degenerate mass case, 
the coefficient of the chiral logarithm is reduced by about $41\%$ (from
-5.5 to -3.25). No dependence upon  $\alpha$ is produced.
 The pattern is opposite for $A_2$ when we compare the quenched
amplitude to the $m_\pi =0$ limit of the full amplitude. In this case
quenching increases the coefficient in absolute value from -0.75 to -3.25,
producing an enhancement of the $\Delta I=3/2$ amplitude.

This 
analysis shows that in most of the cases we have considered here, 
quenching tends to produce sizable modifications
of the coefficients of the chiral
logarithms in the $\Delta I=3/2$ and $\Delta I=1/2$ amplitudes. However,
the knowledge of the parameter $\alpha$ is essential in order to improve
the determination of the  quenched $\Delta I=1/2$ amplitudes.
The comparison with the $m_\pi =0$ limit of the full amplitudes,
expected to be the most approximate to the physical value, has 
shown that the modification induced by quenching (with $\alpha=0$ and 
degenerate quark masses) follows
a pattern that tends to suppress the $\Delta I=1/2$ dominance.

\renewcommand{\theequation}{\arabic{section}.\arabic{equation}}
\setcounter{equation}{0}
\section{Summary and conclusions}
\label{CONC}

We presented the derivation of the ultraviolet divergences of the generating
functional for hadronic weak interactions at one loop in the full theory 
and in the quenched approximation. Weak interactions with $\Delta S=1$,
$\Delta I =1/2$ and  $\Delta I =3/2$, and $\Delta S=2$ have been considered.
The aim was double. Within the full theory we added a few results to the
previous analyses:  we derived the modifications
induced by the inclusion of a dynamical singlet field, and constructed a
minimal basis of divergent counterterms at order $p^4$, both in the octet and 
 27-plet sector. The whole derivation is done for a generic number of 
flavours $N$. 
The list of ultraviolet divergent counterterms and the coefficients of the
divergences in the full theory with $N$ generic are given in Tables
(\ref{TABCTR8}) and (\ref{TABCTR27}).
The generating functional in the full theory and for $N=3$ was already
derived in
Ref. \cite{KMW}, while a minimal basis for the octet sector 
was optimized in Ref. \cite{EKW}; there, the modifications induced by a
singlet dynamical field were anticipated (see appendix A of Ref. \cite{EKW}).
Here, the modifications induced in the generating
functional by the singlet dynamical field  are analyzed in more detail (see
Section (\ref{BOSCR}) for a summary of the results). 

Within the quenched approximation the aim was 
to develop a systematic tool for treating the one loop corrections
of quenched ChPT \cite{qCHPT} in the presence of weak interactions.
The method used here (i.e. the generating functional approach) 
was already proposed in Ref. \cite{npb} in the case of the quenched
approximation of ChPT for strong interactions, while it was 
first introduced in standard ChPT in Refs. \cite{gl84,gl85}.
The presence of weak interactions does not spoil any of the needed 
properties of the generating functional, while we perform an expansion in
powers of the coupling $G_F$.

The cancellation of the flavour number dependence induced by the quenching
procedure is verified within the present approach. The linear flavour number
dependence is cancelled by the fermionic ghost determinant, while the
cancellation of the inverse powers $1/N,\,1/N^2$ is due to the appearance of a
dynamical singlet field. The modifications induced by quenching are summarized
in Tables (\ref{TABCTR8}) and (\ref{TABCTR27}). Quenching largely affects the
structure of the divergences in the octet sector of the theory, where the
number of divergent counterterms goes from 25 to 10, while all the
divergences in the 27-plet sector remain, loosing their flavour number
dependence. 

An interesting feature of the quenched approximation in the presence of weak
interactions is the appearance of new {\em quenched chiral logarithms}
(i.e. of the type $m_0^2\log m_\pi^2$, where $m_0$ is the singlet mass), in
addition to the ones generated by quenched strong interactions.
In Ref. \cite{npb} it 
was shown that quenched chiral logarithms appearing at one 
loop of the strong interactions amount to a redefinition of the 
$B_0$ parameter which measures the quark
condensate in ChPT. Here we have shown that 
the quenched chiral logarithms induced by weak interactions 
can be accounted for via a redefinition of the weak mass term coupling 
$g_8^\prime$ of the leading order weak Lagrangian.

Once the ultraviolet divergences of the weak generating functional to one
 loop are known in the full and in the quenched theory and for degenerate 
quark masses, all the coefficients
 of chiral logarithms are known in both theories, for any nonleptonic weak
 matrix element.
As an application of this result, we have considered
 $B_K$ and $K\to\pi\pi$ matrix elements to
one loop, in order to extract the modifications induced by quenching in the
coefficient of chiral logarithms. At this stage the analysis is done at
 infinite volume. The well known result for $B_K$ \cite{Sharpe}
is found and the structure of the divergent counterterms is clarified for this
quantity. 
For the contribution of chiral logarithms to
$K\to\pi\pi$ matrix elements in the full theory we used the results derived in
Ref. \cite{BPP} and generalized them to a generic number of flavours $N$.
The quenched expression of each matrix element is also given for degenerate
 light quark masses. In Ref. \cite{E2} we shall report a
more detailed numerical comparison of the physical amplitudes with lattice
amplitudes, i.e. where an {\em unphysical} choice of the kinematics is done
with or without the quenched approximation.

Main result of the numerical analysis is that, in most of the cases considered 
here,
the quenched approximation in the degenerate mass case tends to produce 
sizable modifications of  the
coefficients of chiral logarithms in the  $\Delta I=1/2$  and $\Delta
I=3/2$ amplitudes (see Table (\ref{QFULL})).
 However, the knowledge of the singlet parameter $\alpha$ is essential in
 order to improve
the determination of the  quenched $\Delta I=1/2$ amplitudes.

Also, the pattern of the corrections in respect to the $\Delta I=1/2$ rule
goes as follows. 
The contribution from chiral logarithms in the full (unquenched) theory and 
for $m_\pi =0$ 
 clearly produces an enhancement of the $\Delta I=1/2$ amplitude.
 Instead, the comparison of the values of chiral logarithms in the
quenched amplitudes 
with the $m_\pi =0$ values of the chiral logarithms in the 
full amplitudes shows that the
modifications induced by quenching follow a pattern that tends to suppress 
the  $\Delta I=1/2$  dominance.

\section*{Acknowledgements}
The author thanks 
J. Bijnens and J. Prades for useful discussions and collaboration in previous
 works, G. Colangelo for the nice collaboration in recent works of
which this paper is a continuation and
 M. Golterman for providing informations on his related work.
This work has been supported by Schweizerisches Nationalfonds. 
The author acknowledges partial support from the EEC-TMR Program, Contract N.
CT98-0169.

\newpage
\appendix
\renewcommand{\theequation}{\Alph{section}.\arabic{equation}}
\setcounter{equation}{0}

\section{List of weak operators}
\label{LIST}

We define here the weak operators that carry the ultraviolet divergences of the
 weak generating
functional to one loop at order $G_F$. 
With $W_i$ we define the octet operators which are nonzero in the absence of
the singlet field (for them we follow the same ordering as given in Ref.
\cite{EKW}). $D_i,\, i=1,\,\ldots 24$ are 27-plet operators which are nonzero
in the absence of the singlet field. Finally, with $\bar{W}_i,\, i=1,\,
\ldots 23$
and $\bar{D}_i,\, i=1,\,\ldots 3$  we define the extra octet and 27-plet 
operators that are nonzero in the presence of a dynamical singlet field.

The divergent octet $W_i$ operators are defined as follows (the factor $k$ is
factorized everywhere):
\ba
W_1&&\langle \Delta u^2u^2\rangle \nonumber\\
W_2&&\langle \Delta u_\mu u^2u^\mu\rangle \nonumber\\
W_4&&\langle \Delta u_\mu\rangle\langle u^2u^\mu\rangle \nonumber\\
W_5&&\langle \Delta\{\chi_+,u^2\}\rangle \nonumber\\
W_6&&\langle \Delta u_\mu\rangle\langle \chi_+u^\mu\rangle \nonumber\\
W_7&&\langle \Delta\chi_+\rangle\langle  u^2\rangle \nonumber\\
W_8&&\langle \Delta u^2\rangle\langle  \chi_+\rangle \nonumber\\
W_9&&\langle \Delta [\chi_-,u^2 ]\rangle \nonumber\\
W_{10}&&\langle \Delta\chi_+^2\rangle \nonumber\\
W_{11}&&\langle \Delta\chi_+\rangle \langle\chi_+\rangle \nonumber\\
W_{12}&&\langle \Delta\chi_-^2\rangle \nonumber\\
W_{14}&&i\langle \Delta\{f^{\mu\nu}_+,u_\mu u_\nu\}\rangle \nonumber\\
W_{15}&&i\langle \Delta u_\mu f^{\mu\nu}_+ u_\nu\rangle \nonumber\\
W_{16}&&i\langle \Delta\{f^{\mu\nu}_-,u_\mu u_\nu\}\rangle \nonumber\\
W_{18}&&\langle \Delta (f^2_+ -f_-^2)\rangle \nonumber\\
W_{19}&&i\langle \hat{\nabla}_\mu\Delta [u^\mu ,u^2]\rangle \nonumber\\
W_{20}&&\langle \hat{\nabla}_\mu\Delta\{ w^{\mu\nu} ,u_\nu\}\rangle \nonumber\\
W_{21}&&i\langle \hat{\nabla}_\mu\Delta [\chi_+,u^\mu ]\rangle \nonumber\\
W_{22}&&\langle \hat{\nabla}_\mu\Delta d^\mu\chi_+\rangle \nonumber\\
W_{25}&&\langle \hat{\nabla}_\mu\Delta \{f^{\mu\nu}_+,u_\nu\}\rangle 
\nonumber\\
W_{26}&&\langle \hat{\nabla}_\mu\Delta \{f^{\mu\nu}_-,u_\nu\}\rangle 
\nonumber\\
W_{27}&&\langle \Delta (2f^2_+ -\{ f_+,f_-\})\rangle \nonumber\\
W_{36}&&\langle \Delta ([\chi_+,\chi_-]+\chi_+^2-\chi_-^2)\rangle \nonumber\\
W_{37}&&\langle \Delta (f_+ +f_-)^2\rangle \nonumber\\
W_{38}&&\langle \Delta u_\mu\chi_+u^\mu\rangle \, .
\ea
The $\overline{W}_i$ octet operators are (the factor $k$ is
factorized everywhere):
\ba
\bar{W}_1&& \langle\Delta\{u_\mu , u^2\}\rangle\langle u^\mu\rangle \nonumber\\
\bar{W}_2&& \langle\Delta u_\mu u_\nu u^\mu\rangle\langle u^\nu\rangle
 \nonumber\\
\bar{W}_3&& \langle\Delta\{\chi_+, u_\mu\}\rangle\langle u^\mu\rangle 
\nonumber\\
\bar{W}_4&& \langle\Delta [\chi_-, u_\mu ]\rangle\langle u^\mu\rangle 
\nonumber\\
\bar{W}_5&& i\langle\Delta [f^{\mu\nu}_+, u_\mu ]\rangle\langle u_\nu\rangle 
\nonumber\\
\bar{W}_6&&\langle\Delta  u_\nu\rangle\langle f^{\mu\nu}_+\rangle
\nonumber\\
\bar{W}_7&&i\langle\Delta [ u_\mu , u_\nu ]\rangle\langle f^{\mu\nu}_+\rangle
\nonumber\\
\bar{W}_8&& i\langle\Delta [ u_\mu , u_\nu ]\rangle\langle f^{\mu\nu}_-\rangle 
\nonumber\\
\bar{W}_9&&\langle\Delta  f_{\mu\nu +}\rangle\langle f^{\mu\nu}_-\rangle
\nonumber\\
\bar{W}_{10}&&\langle\Delta  f_{\mu\nu -}\rangle\langle f^{\mu\nu}_+\rangle
\nonumber\\
\bar{W}_{11}&& i\langle \hat{\nabla}_\mu\Delta [ u^\mu , u^\nu ]\rangle
\langle u_\nu\rangle \nonumber\\
\bar{W}_{12}&&\langle \hat{\nabla}_\mu\Delta  u_\nu\rangle\langle w^{\mu\nu}
\rangle
\nonumber\\
\bar{W}_{13}&& \langle \hat{\nabla}_\mu\Delta w^{\mu\nu}\rangle\langle
u_\nu\rangle \nonumber\\
\bar{W}_{14}&&  \langle \hat{\nabla}_\mu\Delta  u_\nu\rangle\langle
f^{\mu\nu}_+\rangle 
\nonumber\\
\bar{W}_{15}&& \langle \hat{\nabla}_\mu\Delta f^{\mu\nu}_+\rangle\langle
u_\nu\rangle \nonumber\\
\bar{W}_{16}&& \langle \hat{\nabla}_\mu\Delta u_\nu \rangle\langle
f^{\mu\nu}_- \rangle \nonumber\\
\bar{W}_{17}&& \langle \hat{\nabla}_\mu\Delta f^{\mu\nu}_-\rangle\langle
u_\nu\rangle \nonumber\\
\bar{W}_{18}&& \langle\Delta u_\mu\rangle\langle u^2\rangle \langle
u^\mu\rangle \nonumber\\
\bar{W}_{19}&&  \langle\Delta u_\mu\rangle\langle u^\mu u^\nu\rangle
\langle u^\nu\rangle \nonumber\\
\bar{W}_{20}&& \langle\Delta u^2\rangle\langle u_\mu\rangle\langle
u^\mu\rangle \nonumber\\
\bar{W}_{21}&& \langle\Delta\{ u_\mu ,u_\nu\}\rangle\langle u^\mu\rangle 
\langle u^\nu\rangle \nonumber\\
\bar{W}_{22}&& \langle\Delta \chi_+\rangle\langle u_\mu\rangle\langle
u^\mu\rangle \nonumber\\
\bar{W}_{23}&& \langle\Delta u_\mu\rangle\langle u^\mu\rangle\langle
\chi_+\rangle  \, .
\ea
The 27-plet $D_i$ operators are given by (the factor $q$ is
factorized everywhere):
\ba
D_1&&\langle \Delta\{u_\mu ,u^2\}\rangle\langle\Delta u^\mu\rangle \nonumber\\ 
D_2&&\langle \Delta u_\mu u_\nu u^\mu\rangle\langle\Delta u^\nu\rangle
 \nonumber\\ 
D_3&&\langle \Delta \{ u_\mu , u_\nu\} \rangle\langle\Delta\{ u^\mu ,
 u^\nu\}\rangle
 \nonumber\\
D_4&&\langle \Delta [ u_\mu , u_\nu ] \rangle\langle\Delta [ u^\mu ,
 u^\nu ]\rangle
 \nonumber\\ 
D_5&&\langle \Delta u^2 \rangle\langle\Delta u^2\rangle
 \nonumber\\ 
D_6&&\langle \Delta u_\mu \rangle\langle \Delta u^\mu \rangle\langle u^2\rangle
 \nonumber\\ 
D_7&&\langle \Delta u_\mu \rangle\langle \Delta u_\nu \rangle\langle 
u^\mu u^\nu\rangle \nonumber\\ 
D_8&&\langle \Delta\{\chi_+,u_\mu\}\rangle\langle\Delta u^\mu\rangle 
\nonumber\\ 
D_9&&\langle \Delta u^2 \rangle\langle\Delta \chi_+\rangle
 \nonumber\\ 
D_{10}&&\langle \Delta u_\mu \rangle\langle \Delta u^\mu \rangle\langle 
\chi_+\rangle \nonumber\\ 
D_{11}&&\langle \Delta [\chi_-,u_\mu ]\rangle\langle\Delta u^\mu\rangle 
\nonumber\\ 
D_{12}&&\langle\Delta \chi_+\rangle\langle\Delta \chi_+\rangle
 \nonumber\\ 
D_{13}&&i\langle\Delta [f^{\mu\nu}_+,u_\mu ]\rangle\langle\Delta u_\nu\rangle
 \nonumber\\
D_{14}&&i\langle\Delta f^{\mu\nu}_+\rangle\langle\Delta [u_\mu ,u_\nu ]\rangle
  \nonumber\\
D_{15}&&i\langle\Delta [u_\mu ,u_\nu ]\rangle\langle\Delta f^{\mu\nu}_-\rangle
 \nonumber\\ 
D_{16}&&\langle\Delta f^{\mu\nu}_+\rangle\langle\Delta f_{\mu\nu -}\rangle
 \nonumber\\ 
D_{17}&&i\langle \hat{\nabla}_\mu\Delta [u^\mu ,u^\nu ]\rangle\langle\Delta 
u_\nu\rangle\nonumber\\
D_{18}&&i\langle \hat{\nabla}_\mu\Delta u_\nu\rangle\langle\Delta  [u^\mu
,u^\nu ]\rangle \nonumber\\
D_{19}&&\langle\Delta w^{\mu\nu}\rangle\langle \hat{\nabla}_\mu\Delta
u_\nu\rangle 
\nonumber\\
D_{20}&&\langle \hat{\nabla}_\mu\Delta w^{\mu\nu}\rangle\langle\Delta 
u_\nu\rangle  \nonumber\\ 
D_{21}&&\langle\Delta f^{\mu\nu}_+\rangle\langle \hat{\nabla}_\mu\Delta
u_\nu\rangle 
\nonumber\\
D_{22}&&\langle \hat{\nabla}_\mu\Delta f^{\mu\nu}_+\rangle\langle\Delta 
u_\nu\rangle\nonumber\\ 
D_{23}&&\langle\Delta f^{\mu\nu}_-\rangle\langle \hat{\nabla}_\mu\Delta
u_\nu\rangle 
\nonumber\\
D_{24}&&\langle \hat{\nabla}_\mu\Delta f^{\mu\nu}_-\rangle\langle\Delta 
u_\nu\rangle  \, .
\ea
The 27-plet $\bar{D}_i$ operators are given by (the factor $q$ is
factorized everywhere):
\ba
\bar{D}_1&&\langle \Delta u^2 \rangle\langle \Delta u^\mu \rangle\langle u_\mu
\rangle \nonumber\\ 
\bar{D}_2&&\langle \Delta\{u^\mu ,u^\nu\}\rangle\langle\Delta u_\mu\rangle
\langle u_\nu\rangle  \nonumber\\ 
\bar{D}_3&&\langle \Delta u^\mu\rangle\langle \Delta\chi_+ \rangle\langle u_\mu
\rangle  \, .
\ea
The projection operator $\Delta_{ij}$ is defined as
$ \Delta_{ij}=u \lambda_{ij}u^\dagger$, with
$\left (\lambda_{ij}\right )_{ab} = \delta_{ia}\delta_{jb}$.
The building blocks used in the definition of the counterterms are as follows 
\ba
u_\mu&=& iu^\dagger D_\mu U u^\dagger = u_\mu^\dagger  \nonumber\\
\chi_\pm&=& u^\dagger \chi u^\dagger\pm u\chi^\dagger u  \nonumber\\
f^{\mu\nu}_\pm &=& u f^{\mu\nu}_L u^\dagger \pm u^\dagger f^{\mu\nu}_R u
 \nonumber\\
w_{\mu\nu}&=& d_\mu u_\nu + d_\nu u_\mu\, .
\label{BB}
\ea
Also, the relations $f^{\mu\nu}_-=  d^\nu u^\mu - d^\mu u^\nu$ 
and  $f^{\mu\nu}_+=2i\Gamma^{\mu\nu} -i/2 [u^\mu ,u^\nu ]$
are useful.
We defined as in Ref. \cite{EKW} the covariant derivative acting on
$\Delta$ as $\hat{\nabla}_\mu \Delta = d_\mu\Delta +{i\over 2} [u_\mu ,\Delta
]$. The covariant derivative $d_\mu$ is the usual one acting on the fields in
Eq. (\ref{BB}). It is defined as 
\ba
d_\mu O&=& \partial_\mu O + [\Gamma_\mu ,O]\, , \nonumber\\
\Gamma_\mu &=& {1\over 2} ( [u^\dagger ,\partial_\mu u ]-i u^\dagger
r_\mu u -i u l_\mu u^\dagger )\, .
\ea
The equation of motion relates the field $\chi_-$ to the covariant derivative
of the $u_\mu$ field
\be
w_{\mu\mu} -{1\over N}\langle w_{\mu\mu}\rangle{\bf 1} = i\left (
\chi_- - {1\over N}\langle\chi_-\rangle{\bf 1}\right )\, .
\ee

\section{Integral over the singlet fields}
\label{SINGLET}

In this Appendix we give the explicit expressions of the single terms that
contribute to the ultraviolet divergent part 
in the perturbative expansion of the logarithm of the determinant
of the singlet operator as defined in Eq. (\ref{UVEXP}). The result for each
 term reads as follows 
\ba
&&\hspace{-1.5truecm}{i\over 2} {\mbox{Tr}} D_X^{0-1} A_w =\, 
-{1\over (4\pi )^2(d-4)}\,\int~dx\, k \left\{
{1\over 2N}(g_8^\prime -g_8)\left (1-{N\over 3}\alpha\right ) M^2
\langle\Delta\chi_+\rangle \right. \nonumber\\
&&\hspace{-1.5truecm}\left.  -{\bar{g}_8\over 4}M^2 \langle\Delta\chi_+\rangle
+{1\over 6}(g_8^\prime -g_8)(m_0^2-\alpha M^2) \langle\Delta\chi_+\rangle  
\right\}+O(G_F^2)\, ,
\ea
\ba
&&\hspace{-1.5truecm}-{i\over 2} {\mbox{Tr}} D_X^{0-1} A_s D_X^{0-1} A_w =\,
-{1\over (4\pi )^2(d-4)}\,\int~dx \,  k \left\{ 
{\alpha^2\over 36}(g_8^\prime -g_8)
\langle\Delta\chi_+\rangle\langle\hat{\chi}_+\rangle  \right.
 \nonumber\\
&&\hspace{-1.5truecm}
+ {1\over 4N^2}(g_8^\prime -g_8)\left (1-{2\over 3}N\alpha\right )
 \langle\Delta\chi_+\rangle\langle\hat{\chi}_+\rangle
- {\bar{g}_8 \over 8N}\left (1-{N\over 3}\alpha\right )
 \langle\Delta\chi_+\rangle\langle\hat{\chi}_+\rangle \nonumber\\
&&\hspace{-1.5truecm} -{1\over 2}\langle\hat{\chi}_+\rangle 
\left (\tilde{v}_8 k\langle\Delta u^2\rangle 
+\tilde{v}_5 k\langle\Delta\chi_+\rangle
+\tilde{v}_{27}q\langle\Delta
u_\mu\rangle \langle\Delta u^\mu\rangle +\tilde{v}_0 k
\langle\Delta u_\mu\rangle \langle u^\mu\rangle \right ) \nonumber\\
&&\hspace{-1.5truecm}\left.
 -{1\over 2}(g_8^\prime -g_8) \langle\Delta\chi_+\rangle 
\left ( v_1\langle u^2\rangle +v_2 \langle\hat{\chi}_+\rangle \right )
\right\}+O(G_F^2)\, ,
\ea
\ba
&&\hspace{-1.5truecm}-{i\over 2} {\mbox{Tr}} ( D_X^{0-1}B_s^T \bar{D}^{-1}B_w +
 D_X^{0-1}B_w^T \bar{D}^{-1}B_s  ) =\, -{1\over (4\pi )^2(d-4)}\,\int~dx
  \nonumber\\ 
&&\hspace{-1.5truecm}
\left\{ -{g_8\over 6}(m_0^2-\alpha M^2)k \langle\Delta\chi_+\rangle   
+ g_8 k\left (1-{N\over 3}\alpha\right )
\left [
-{i\over 4N}\langle [d_\mu +{i\over 2} u_\mu ,[\Delta , u^\mu ]]\chi_+\rangle
\right. \right.\nonumber\\
&&\hspace{-1.5truecm}\left.  -{1\over 2N}\langle d^2\Delta\chi_+\rangle
-{1\over 2N}\langle\Delta\chi_+^2\rangle
+{1\over 2N^2}\langle\Delta\chi_+\rangle\langle\chi_+\rangle
\right ] \nonumber\\
&&\hspace{-1.5truecm} + g_8^\prime k\left (1-{N\over 3}\alpha\right )\left [
{1\over 2N}\langle\Delta\chi_+^2\rangle -
{1\over 2N^2}\langle\Delta\chi_+\rangle\langle\chi_+\rangle
+{1\over 4N}\langle\Delta [\chi_+, \chi_-]\rangle
\right ]
 \nonumber\\ 
&&\hspace{-1.5truecm}
-{1\over 4N} g_{27} \left (1-{N\over 3}\alpha\right ) q
\langle\Delta\chi_+\rangle \langle\Delta\chi_+\rangle   
+{1\over 2}g_8 k \langle\Delta\chi_+\rangle 
\left ( v_1\langle u^2\rangle +v_2 \langle\hat{\chi}_+\rangle \right )
\nonumber\\ 
&&\hspace{-1.5truecm} +\bar{g}_8 k\left [
{\alpha\over 24}\langle\Delta\chi_+\rangle\langle\hat{\chi}_+\rangle
-{i\over 8}\langle [d_\mu +{i\over 2}u_\mu ,[\Delta , u^\mu ]]\chi_+\rangle
 -{1\over 4}\langle d^2 \Delta\chi_+\rangle
-{1\over 8}\langle\Delta\chi_+^2\rangle      \right ]  \nonumber\\ 
&&\hspace{-1.5truecm}\left.
  -g_8 k{\alpha\over 12}\left [ \langle\Delta u_\mu\chi_+ u^\mu\rangle  
-{1\over 2}\langle\Delta\{ \chi_+,u^2\}\rangle 
- \langle\Delta\chi_+^2\rangle +{1\over N}
  \langle\Delta\chi_+\rangle \langle\chi_+\rangle \right ]
\right\} +O(G_F^2)\, ,\nonumber\\
&&
\ea
where $\bar{D}^{-1}$ is expanded up to order $G_0^2$ when the term
proportional to $\alpha \Box$ inside $B_w$ is considered.
\ba
&&\hspace{-1.5truecm}
{i\over 2}{\mbox{Tr}}\left[{D_X^0}^{-1}A_s {D_X^0}^{-1}\left
    (B_s^T\overline{D}^{-1}B_w +B_w^T\overline{D}^{-1}B_s\right )\right ]
=  -{1\over (4\pi )^2(d-4)}\,\int~dx\, \nonumber\\ 
&&\hspace{-1.5truecm}  g_8 {\alpha\over 12 N}  \langle\Delta\chi_+\rangle 
\langle\hat{\chi}_+\rangle  +O(G_F^2)\, .
\label{B11}
\ea
In Eq. (\ref{B11})
only one term gives contribution to the ultraviolet divergent part of the
singlet determinant.
We used $u^2\equiv u_\mu u^\mu$ and $\langle\hat{\chi}_+\rangle = 
\langle\chi_+\rangle -2NM^2$.
In the derivation of the complete result in terms of the operators
listed in Appendix \ref{LIST} some relations are useful
\ba
i\langle [d_\mu +{i\over 2} u_\mu , [\Delta ,u^\mu ]]\chi_+\rangle&=&
-i\langle\hat{\nabla}_\mu \Delta [\chi_+,u^\mu ]\rangle +{1\over 2}
\langle\Delta [\chi_+ ,\chi_- ]\rangle \nonumber\\ 
\langle d^2 \Delta\chi_+\rangle&=& -\langle\hat{\nabla}_\mu \Delta d^\mu\chi_+
\rangle +{1\over 4}\langle\Delta [\chi_+ ,\chi_- ]\rangle 
-{i\over 2} \langle\hat{\nabla}_\mu \Delta [\chi_+,u^\mu ]\rangle 
 \nonumber\\
&&-{1\over 4}\langle\Delta\{\chi_+, u^2\} \rangle
+{1\over 2}\langle\Delta  u_\mu\chi_+u^\mu\rangle\, ,
\ea
where $\hat{\nabla}_\mu \Delta = d_\mu\Delta +{i\over 2} [u_\mu ,\Delta
]$ and the building blocks are defined in Eq. (\ref{BB}).

\end{document}